\documentclass[aps,prd,superscriptaddress,nofootinbib,tighten,preprint]{revtex4}
\usepackage[utf8]{inputenc}
\usepackage[english]{babel}
\usepackage{amsmath}
\usepackage{subfigure}
\usepackage{color}
\usepackage{amsfonts}
\usepackage{amssymb}
\usepackage{graphicx}
\newcommand{\be}{\begin{equation}}
\newcommand{\ee}{\end{equation}}
\newcommand{\bea}{\begin{eqnarray}}
\newcommand{\eea}{\end{eqnarray}}

\catcode`@=12

\begin{document}

\title{Axion Dark Matter and additional BSM aspects in an extended 2HDM setup}
\author{Giorgio Arcadi}
\email{giorgio.arcadi@unime.it}
\affiliation{ 
Dipartimento di Scienze Matematiche e Informatiche, Scienze Fisiche e Scienze della Terra, Universita degli Studi di Messina, Via Ferdinando Stagno d'Alcontres 31, I-98166 Messina, Italy
}
\affiliation{INFN Sezione di Catania, Via Santa Sofia 64, I-95123 Catania, Italy.}

\author{Sarif Khan}
\email{sarif.khan@uni-goettingen.de}
\affiliation{Institute for Theoretical Physics, 
Georg-August University G\"ottingen,
Friedrich-Hund-Platz 1, G\"ottingen, D-37077 Germany}

\begin{abstract} 
We illustrate and discuss the phenomenology of a model featuring a two-Higgs doublet sector augmented by two $SU(2)$ singlet scalars. The gauge symmetry group is extended as well with a $U(1)_{B_{i}-L_{i}}$ component whose spontaneous breaking leads to the gauge boson which has an important effect in the muon (g-2). A global
PQ symmetry is introduced upon its breaking we have the axion particle which is also
DM in our work. In particular, we have focussed on Type-X and Type-II 2HDM models and found out that
(g-2) can not be explained only by the scalar sector for Type-II 2HDM mainly due to
stringent constraint from $b \rightarrow s \gamma$ resulted in $M_{H^{\pm}} > 800$ GeV.
For Type-II 2HDM, we can have axion coupling with the gluons which generates the axion
potential and possible explanation for the strong CP problem.
The proposed model accommodates neutrino masses via the Type-I see-saw mechanism with an upper bound on the right-handed neutrino mass 1 GeV (1 TeV) for Type-II (Type-X) 2HDM due to the presence of 
Planck scale 
suppressed operators. Moreover, we also have additional scalars which affect the 
oblique parameters and hence the W-boson mass which leads us to explain the W-boson mass observed at
CDF-II detector. The most stringent constraints on the masses and quartic couplings come 
from the perturbativity and potential bound from below conditions which leads
to fine-tuning among the parameters in part of the parameter space. 
Finally, we discuss the possible detection prospects of the axion DM 
and the additional gauge boson.
\end{abstract}
\maketitle
\section{Introduction}
\label{Intro}

In the present work, we aim to address several drawbacks which are intact in pure Standard Model (SM) 
and demand beyond SM physics to tackle them. With the progression of experiments, few important
limitations within the SM have come to light, including the presence of dark 
matter (DM) \cite{Bertone:2016nfn}, 
non-zero neutrino mass \cite{Kajita:2016cak, McDonald:2016ixn, Esteban:2020cvm}, 
disagreement between the theoretical and experimental values of 
the muon ($g-2$) \cite{Aoyama:2020ynm}, 
and larger values of W-boson mass as confirmed by the CDF-II measurement \cite{CDF:2022hxs}.
Confirmation of DM's existence has been derived from several observations, notably evident in the flatness of the rotational curve \cite{Sofue:2000jx}, 
discrepancies in the centre of masses of the visible sector compared to the total mass in the bullet cluster data \cite{Clowe:2003tk}, and the precise measurement of 
the DM quantity through cosmic microwave background data collected by 
the Planck satellite \cite{Planck:2018vyg}.
 Axion is one of the most popular and best motivated proposals to address the DM puzzle. The existence of the axion is originally connected with the strong CP problem, namely the suppression of the following topological term in the QCD Lagrangian:
\begin{eqnarray}
    \mathcal{L}^{\theta}_{CP} = \left( \theta + arg\left[ det(m_{q}) \right] \right)
    \frac{g^2_{s}}{32 \pi^2} G_{\mu\nu} \widetilde{G^{\mu\nu}}\,,
\end{eqnarray}
where $g_{s}$ is the strong sector gauge coupling, $m_{q}$ is the quark mass matrix and
$\widetilde{G^{\mu\nu}} = \epsilon^{\mu\nu\rho\sigma} G_{\rho\sigma}$ is the dual of
the strong sector field strength tensor $G_{\mu\nu}$.
The presence of this term explicitly breaks the Parity symmetry, resulting in the generation of a non-zero electric dipole moment for the neutron ($d_n$), estimated around 
$d_{n} \sim 10^{-16} \Bar{\theta} \,\,{\rm e\,cm}$ 
(where $\bar{\theta} = \theta + arg\left[ det(m_{q}) \right]$). However, the current experimental limit, 
$d_{n} < 10^{-26}\,\,{\rm e\,cm}$, imposes a stringent constraint on the $\theta-$term, indicating
$\bar{\theta} < 10^{-10}$ \cite{Baker:2006ts}. The remarkably small magnitude of $\bar{\theta}$ 
seems unnatural; given its nature as an angular variable 
would typically be expected as $\mathcal{O}(1)$. Peccei-Quinn
mechanism \cite{Peccei:1977hh, Peccei:1977ur, Weinberg:1977ma, Wilczek:1977pj} allows to set the $\bar \theta$ parameter via the introduction of an appropriate symmetry, with the axion being a pseudogoldstone boson associated to the latter. Observational constraints favour the axion to have extremely suppressed couplings with the ordinary matter which make it cosmologically stable and hence a potential DM candidate.
Another limitation of the SM is represented by the origin of neutrino masses.
In the SM, no mass terms are present of the neutrinos, as there are no right-handed counterparts to the left-handed neutrinos present in the $SU(2)$ lepton doublets. Oscillation data by many experiments \cite{Super-Kamiokande:1998kpq, SNO:2002tuh, KamLAND:2002uet, DayaBay:2015lja} contradict the SM prediction as they require massive neutrinos, with sub-ev mass splittings, to be accounted for. 
Another problem is the measurement of the muon ($g-2$) which disagrees with the theoretical
value measured by the SM \cite{Aoyama:2020ynm} and 
the experimental observation \cite{Muong-2:2021ojo}. 
The recent measurement of 
muon ($g-2$) at FNAL disagrees with the SM value at $4.2\,\sigma$. It is worth mentioning that
the measurement of vacuum hadronic polarisation contribution by different methods can decrease the
disagreement. For example, the lattice calculation by BMW collaboration \cite{Borsanyi:2020mff} 
has reduced the anomaly down to $1.5\,\sigma$. Moreover, the recent measurement of 
$e^{+}e^{-} \rightarrow \pi^{+}\pi^{-}$ cross-section at the CMD-III experiments \cite{CMD-3:2023alj} 
has also decreased the discrepancy between the theoretical and experimental 
values to $2.4\,\sigma$. However, the lattice data and CMD-III data are not final and they also
need to be confirmed before any conclusion. In this work, we have focussed on explaining
the disagreement between the muon $g-2$ values obtained theoretically from SM and at 
FNAL experiment. 
There is also another drawback of SM in the measurement of the W-boson mass if we 
consider the CDF-II measurement \cite{CDF:2022hxs}. The latter disagrees with the SM value by $7\,\sigma$, hence calling of an extra BSM contribution to the mass of the gauge boson. 
In this work, we propose a model addressing the DM puzzle (via axion), the origin of neutrino masses as well as the observational hints summarized above. 
The model consists of an extended Higgs sector, made by two $SU(2)$-doublets and two complex scalar singlets, as well as an extended gauge group, with a spontaneously broken anomaly-free (via suitable charge assignments of the field) $U(1)_{B_i-L_i}$. The two complex scalar singlet provide additional pseudoscalar degrees of freedom which allow to provide a longitudinal component to the gauge boson of the new symmetry and, at the same time, to incorporate an axion field. Three right-handed neutrinos are, finally, present which allow to generate the SM neutrino masses via see-saw. Analogously to the conventional 2HDM we have assumed ad hoc $Z_2$ symmetries, as well as specific assignations of the charges of the scalars under the new symmetry, to prevent tree-level FCNC \cite{Glashow:1976nt, Paschos:1976ay}. Interestingly this combination of charge assignments and discrete symmetries resembles an accidental PQ symmetry \cite{Peccei:1977hh, Peccei:1977ur, Weinberg:1977ma, Wilczek:1977pj}. This kind of possibility has been explored e.g. in \cite{Bjorkeroth:2017tsz, Harigaya:2013vja, Dias:2002hz, Ahn:2014gva, Honecker:2015ela} 
and can be easily implemented in our work without impact on the phenomenology. Alternatively, one could consider explicitly introducing a global $U(1)_{PQ}$ symmetry \cite{Clarke:2015bea}. As well known, however, global symmetries are expected to be broken by gravity \cite{Hawking:1987mz, Lavrelashvili:1987jg, Giddings:1988cx, Coleman:1988tj, Gilbert:1989nq}.
In the present work due to the charge assignments, we have found that the axion field
and right-handed neutrino masses
originate from the same fields by its CP odd component and vev value, respectively. 
The axion and neutrino
masses in 2HDM set-up have been studied 
in \cite{Clarke:2015bea,Dias:2021lmf} but our work is completely different. 
We have focussed on Type-II and Type-X type 
2HDM model. As we will see later, in explaining the muon ($g-2$) from Type-II 2HDM, we need 
$100\%$ contribution from the additional gauge boson because of the very tight constraint
on the charged Higgs mass $M_{H^{\pm}} > 800$ GeV coming from the $b \rightarrow s \gamma$ 
measurement \cite{HFLAV:2016hnz, Misiak:2017bgg}. On the other hand, for Type-X 2HDM, we can explain the muon $(g-2)$
from the Higgs sector as well in part of the region but the immediate drawback of Type-X 
is that it can not explain the 
strong CP problem in contrast to Type-II 2HDM which can not
generate 100\% muon ($g-2$) only from the Higgs sector. In the case of neutrino mass, it does not
distinguish the Type-X and Type-II 2HDM because the Dirac mass matrix depends on the
Higgs doublet $H_2$
and right-handed neutrino masses depend on singlet scalars.
As will be discussed later, to explain the $(g-2)$ we can not choose the gauge coupling and gauge 
boson mass randomly which will impact the right neutrino mass.
Due to the dependence of a few elements in the right-handed neutrino mass matrix Planck scale,
we can not get the right-handed neutrino masses above 1 GeV (1 TeV) for the Type-II (Type-X). 
This hierarchy in RH neutrino masses for two types of 2HDM ensures the 
 Dirac Yukawa coupling for the neutrinos can reach a maximal value $\mathcal{O}(10^{-8})$ 
 $\left(\mathcal{O}(10^{-6})\right)$ 
 for Type-II (Type-X) 2HDM model. For both Type-II and Type-X, we can explain the 
 W-boson mass observed by the CDF-II detector and also the regions in $S, T, U$ planes 
 which are consistent with SM predicted value of W-boson mass.

 The rest of the paper is organised as follows. In Section \ref{model}, 
 we have explained the model in detail.
 The neutrino mass has been discussed in Section \ref{NM-Type-I} and the axion DM 
 study has been shown in Section \ref{axion-part}. In Section \ref{muon-g-2}, we have
 discussed the muon ($g-2$) followed by the allowed regions in Section \ref{result}.
 The W-boson mass has been studied in Section \ref{w-mass} and finally we have presented 
 our conclusion in Section \ref{conclusion}.

\section{Model}
\label{model}

In the present work, we have tried to explain the SM drawbacks namely the presence of
dark matter, neutrino mass, excess contribution in muon ($g-2$) anomaly, strong CP problem 
and extra contribution to W-boson mass as obtained by 
CDF-II data. In this context, we have tried to explain
all of them from a common origin. In particular, we have considered axion DM in the context 
of the DFSZ type axion model \cite{Zhitnitsky:1980tq, Dine:1981rt}. 
As we know, the CP odd component present in 2HDM models can not 
be axion DM because of many constraints {\it e.g.} supernovae bound 
on axion decay constant\cite{Chang:2018rso}. Therefore, we extended the Higgs sector by 
two additional singlet scalars and an abelian gauge symmetry. 
Among the extra two singlet scalars, one of them
is neutral under the new abelian gauge symmetry and the CP odd component coming from it can be
an axion DM and the other CP odd component which is charged under abelian gauge symmetry becomes 
the longitudinal component of additional gauge boson and impart mass to it. Moreover, 
we have introduced three right-handed neutrinos for generating the neutrino mass and their
masses depend on the vevs of the singlet scalars. 

\begin{center}
\begin{table}[h!]
\begin{tabular}{||c|c|c|c||}
\hline
\hline
\begin{tabular}{c}
    Gauge\\
    Group\\ 
    \hline
    
    ${\rm SU(2)}_{\rm L}$\\ 
    \hline
    ${\rm U(1)}_{Y}$\\ 
\end{tabular}
&

\begin{tabular}{c|c|c}
    \multicolumn{3}{c}{Baryon Fields}\\ 
    \hline
    $Q_{L}^{i}=(u_{L}^{i},d_{L}^{i})^{T}$&$u_{R}^{i}$&$d_{R}^{i}$\\ 
    \hline
    $2$&$1$&$1$\\ 
    \hline
    $1/6$&$2/3$&$-1/3$\\ 
\end{tabular}
&
\begin{tabular}{c|c|c}
    \multicolumn{3}{c}{Lepton Fields}\\
    \hline
    $L_{L}^{i}=(\nu_{L}^{i},e_{L}^{i})^{T}$ & $e_{R}^{i}$ & $N_{i}$\\
    \hline
    $2$&$1$&$1$\\
        \hline
    $-1/2$&$-1$&$0$\\
\end{tabular}
&
\begin{tabular}{c|c|c|c}
    \multicolumn{4}{c}{Scalar Fields}\\
    \hline
    $H_{1}$&$H_{2}$&$\phi_{1}$&$\phi_{2}$\\
    \hline
    $2$&$2$&$1$&$1$\\
    \hline
    $1/2$&$1/2$&$0$&$0$\\
\end{tabular}\\
\hline
\hline
\end{tabular}
\caption{Particle contents and their corresponding
charges under the SM gauge group where $i$ indicates the three generations for quarks and leptons.}
\label{tab1}
\end{table}
\end{center}

\begin{center}
\begin{table}[h!]
\begin{tabular}{||c|c|c|c||}
\hline
\hline
\begin{tabular}{c}
    Gauge\\
    Group\\ 
    
    \hline
    $U(1)_{B_{i} - L_{i}}$\\ 
\end{tabular}
&

\begin{tabular}{c|c|c}
    \multicolumn{3}{c}{Baryon Fields}\\ 
    
\hline
    $Q_{L}^{i}$&$u_{R}^{i}$&$d_{R}^{i}$\\ 
    \hline
    $b_{i}$&$b_{i}$&$b_{i}$\\ 
\end{tabular}
&
\begin{tabular}{c|c|c}
    \multicolumn{3}{c}{Lepton Fields}\\
    \hline
    ( $L_{L}^{e}$, $e_R$, $N_e$ ) & ( $L_{L}^{\mu}$, $\mu_R$, $N_{\mu}$ ) & ( $L_{L}^{\tau}$, $\tau_{R}$, $N_{\tau}$ )
    \\
    \hline
    $b^{\prime}_{1}$&$b^{\prime}_{2}$&$b^{\prime}_{3}$\\
\end{tabular}
&
\begin{tabular}{c|c|c|c}
    \multicolumn{4}{c}{Scalar Fields}\\
    \hline
    $H_{1}$&$H_{2}$&$\phi_{1}$&$\phi_{2}$\\
    \hline
    $0$&$0$&$0$&$1$\\
\end{tabular}\\
\hline
\hline
\end{tabular}
\caption{
Charges of the particles under the new abelian gauge group $U(1)_{B_{i}-L_{i}}$. 
The fermion doublets $L^{i}_{L}$ and $Q^{i}_{L}$ are defined in Table \ref{tab1}.}
\label{tab2}
\end{table}
\end{center}

The particle content of the model, together with the charge assignments of the individual fields, is summarized in Tables [\ref{tab1}, \ref{tab2}]. In the tables $H_{1,2}$ stand for the two $SU(2)$ 
doublets while $\phi_{1,2}$ are the new singlets.
The additional $U(1)_{B_{i}-L_{i}}$ represent any combination  
of the baryonic and leptonic charges of the fields which must
obey the following gauge anomaly condition as described in Section \ref{gauge-anomaly}
of appendix,
\begin{eqnarray}
    \sum^{3}_{i=1} \left( 3 b_{i} + b^{\prime}_{i} \right) = 0\,.
\end{eqnarray}
As can be seen from the above equation by obeying the relation we can have many 
combinations for $b_{i}, b^{\prime}_{i}$ ($i = 1,2,3$). In the present work, we have chosen 
$b_{i} = 0$, $b^{\prime}_{1} = 0$ and $b^{\prime}_{2} = - b^{\prime}_{3} = 1$ which 
implies $(B_{i}-L_{i}) = L_{\mu} - L_{\tau}$. As will be clarified below, this specific choice for the abelian group is crucial to account for the generation of neutrino masses and to explain the $g-2$ anomaly. Furthermore, it can provide an accidental  
continuous global symmetry depending on the introduction of additional discrete
symmetries.
The model discussed in this work is described by the following Lagrangian: 
\begin{eqnarray}
\mathcal{L}&=&\mathcal{L}_{2HDM} + \mathcal{L}_{Y} + \mathcal{L}_{N}
+ \sum_{i = 1,2} (D_{\mu}\phi_{i})^{\dagger} (D^{\mu}\phi_{i})
-\mathcal{V}(H_{1},H_{2},\phi_{1},\phi_{2})
-\frac{1}{4} F_{\mu \tau}^{i\,\alpha \beta} {F^{i}_{\mu \tau}}_{\alpha \beta}
\,, 
\label{lag}
\end{eqnarray}
where $\mathcal{L}_{2HDM}$ consists of kinetic terms
for all the fields for pure 2HDM model \cite{Branco:2011iw}, $\mathcal{L}_{Y}$ is the Yukawa terms 
associated with the SM quarks and leptons,
\begin{eqnarray}
\mathcal{L}_{Y} = y^{u}_{ij} \bar{Q}^i_{L} \tilde{H}_2 u^j_{R} + 
y^{d}_{ij} \bar{Q}^i_{L} H_{m} d^j_{R} + 
y^{l}_{ij} \bar{L}^i_{L} H_{n} e^j_{R} + {\it h.c.}\,.
\end{eqnarray}
where $\tilde{\phi}_2 = i \sigma_{2} \phi^{*}_{2}$ and $m,n = 1,2$. Depending on $m,n=1,2$
we can have four kinds of scenarios which are mentioned in Table \ref{tab3} with the associated 
Peccei-Quinn charges for the quarks, leptons and scalars. 
As pointed out in \cite{Glashow:1976nt, Paschos:1976ay}, the tree level
flavour changing neutral current (FCNC) can be forbidden if the same charge fermions get mass from the 
one scalar doublet. With this freedom, we can have four different types of combination
for 2HDM model referred to as the Type-I, II, X and Y depending on the $H_{1,2}$
interaction with the quarks and leptons (shown in Table \ref{tab3}). The four different types of 2HDM
model can be
achieved by introducing an additional global 
$U(1)_{PQ}$ symmetry \cite{Peccei:1977hh, Peccei:1977ur, Weinberg:1977ma, Wilczek:1977pj} with the charge 
assignment as shown in Table \ref{tab3} and the same global symmetry will also help us in solving the 
strong CP problem. 
In this setup, the $U(1)_{PQ}$ charges 
shown in Table \ref{tab3} can introduce additional PQ-violating terms at the Planck scale which 
may shift the axion potential because the global symmetry is not conserved at the Planck scale 
\cite{Hawking:1987mz, Lavrelashvili:1987jg, Giddings:1988cx, Coleman:1988tj, Gilbert:1989nq}.
This kind of situation can be evaded by introducing additional discrete symmetries which 
will make the PQ symmetry as the accidental global symmetry 
\cite{Bjorkeroth:2017tsz, Harigaya:2013vja, Dias:2002hz, Ahn:2014gva, Honecker:2015ela} 
and will forbid the extra PQ
breaking terms at the Planck scale or will have a minuscule impact at the axion potential.
Among the four choices of the 2HDM model, we will see in the later part
that Type-II and Type-Y 2HDM models can address the strong CP problem  
and the corresponding CP-odd scalar will become the axion DM. 
The other two choices Type-I, X can not solve the strong CP problem but 
still can produce axion-type DM by 
the misalignment mechanism \cite{Preskill:1982cy, Abbott:1982af}.
\begin{center}
\begin{table}[h!]
\begin{tabular}{||c|c|c|c|c||}
\hline
\hline
\begin{tabular}{c}
    Global\\
    Symmetry\\ 
    
    \hline
    \\
    $U(1)_{PQ}$\\ 
    \\
    \\
\end{tabular}
&
\begin{tabular}{c|c}
    \multicolumn{2}{c}{Classification}\\
    \hline
    Type &$(m,n)$
    \\
    \hline
    I&$(2,2)$\\
        \hline
    II&$(1,1)$\\
            \hline
    X&$(2,1)$\\
                \hline
    Y&$(1,2)$\\
\end{tabular}
&
\begin{tabular}{c|c|c|c}
    \multicolumn{4}{c}{Scalar Fields}\\
    \hline
    $H_{2}$&$H_{1}$&$\phi_{1}$&$\phi_{2}$\\
    \hline
    $\cos^{2}\beta$&$-\sin^{2}\beta$&$-1$&$0$\\
        \hline
    $\cos^{2}\beta$&$-\sin^{2}\beta$&$-1$&$0$\\
            \hline
    $\cos^{2}\beta$&$-\sin^{2}\beta$&$-1$&$0$\\
                \hline
    $\cos^{2}\beta$&$-\sin^{2}\beta$&$-1$&$0$\\
\end{tabular}
&

\begin{tabular}{c|c|c}
    \multicolumn{3}{c}{Baryon Fields}\\ 
    
\hline
    $Q_{L}^{i}$&$u_{R}^{i}$&$d_{R}^{i}$\\ 
    \hline
    $X_{u}$&$X_{u} +\cos^{2}\beta$&$X_{u} -\cos^{2}\beta$\\
        \hline
    $X_{u}$&$X_{u} +\cos^{2}\beta$&$X_{u} +\sin^{2}\beta$\\
            \hline
    $X_{u}$&$X_{u} +\cos^{2}\beta$&$X_{u} -\cos^{2}\beta$\\
                \hline
    $X_{u}$&$X_{u} +\cos^{2}\beta$&$X_{u} +\sin^{2}\beta$\\
\end{tabular}
&
\begin{tabular}{c|c|c}
    \multicolumn{3}{c}{Lepton Fields}\\
    \hline
    $L_{L}^{j}$ &  $e^{j}_{R}$ &  $N^{j}_{R}$
    \\
    \hline
    $X_{l}$&$X_{l} -\cos^{2}\beta$ &$X_{l} +\cos^{2}\beta$\\
        \hline
    $X_{l}$&$X_{l} +\sin^{2}\beta$ &$X_{l} +\cos^{2}\beta$\\
            \hline
    $X_{l}$&$X_{l} +\sin^{2}\beta$ &$X_{l} +\cos^{2}\beta$\\
                \hline
    $X_{l}$&$X_{l} -\cos^{2}\beta$ &$X_{l} +\cos^{2}\beta$\\
\end{tabular}
\\
\hline
\hline
\end{tabular}
\caption{
PQ charges of the particles under Peccei-Quinn symmetry $U(1)_{PQ}$. The charges of 
$L^{i}_{L}$ and $Q^{i}_{L}$ are free parameters as defined by $X_{l}$ and $X_{u}$,
respectively, and can be chosen according to the purpose. Additionally, adhering to the conservation of global PQ symmetry at the Planck scale, the relationship between $X_{l}$ and $\beta$ within the neutrino sector Lagrangian, expressed as Eq. \ref{lagN}, is defined as $X_{l} = \frac{1}{2} - \cos^{2}\beta$.}
\label{tab3}
\end{table}
\end{center}
The Lagrangian associated with the right-handed neutrinos 
takes the following form after obeying all the gauge symmetry and PQ symmetry,
\begin{eqnarray}
\mathcal{L}_{N}&=&
\sum_{i=e,\mu,\tau}\frac{i}{2}\bar{N^i_L}\gamma^{\mu}D^{N}_{\mu} N^{i}_L
-\sum_{i=e,\,\mu,\,\tau} y_{ii} \bar{L_{i}}
\tilde{H_{2}} N_{i} 
- M_{ee} N_{e} N_{e} \frac{\phi_{1}}{M_{pl}} - \lambda_{e\mu} N_{e} N_{\mu} \phi_2 \frac{\phi_{1}}{M_{pl}}
 \nonumber \\
&&
- M_{e\tau} N_{e} N_{\tau} \frac{\phi_{1}}{M_{pl}} 
- \lambda_{\mu\tau} N_{e} N_{\tau} \phi_2 \frac{\phi_{1}}{M_{pl}} 
+h.c.\,
\label{lagN}
\end{eqnarray}

The potential consists of all the scalar fields after respecting all
the symmetries in consideration have the following form,
\begin{eqnarray}
\mathcal{V}(H_{1},H_{2},\phi_1, \phi_2) &=& 
- \mu^2_{11} (H^{\dagger}_{1} H_{1}) - \mu^2_{22} (H^{\dagger}_{2} H_{2})
- \mu^2_{\phi_1} (\phi^{\dagger}_1 \phi_{1}) - \mu^2_{\phi_2} (\phi^{\dagger}_2 \phi_{2}) 
+ \mu \left( (H^{\dagger}_{1} H_{2}) \phi_1 + {\it h.c.} \right)
\nonumber \\
  && + \lambda_{1} (H^{\dagger}_{1} H_{1})^{2} + 
  \lambda_{2} (H^{\dagger}_{2} H_{2})^{2} + \lambda_{\phi_1} 
  (\phi^{\dagger}_{1} \phi_1)^2 + \lambda_{\phi_2} 
  (\phi^{\dagger}_{2} \phi_2)^2 
  + \lambda_{12} (H^{\dagger}_{1} H_{1}) (H^{\dagger}_{2} H_{2})
  \nonumber \\
  &&
    + \lambda^{\prime}_{12} (H^{\dagger}_{1} H_{2}) (H^{\dagger}_{2} H_{1})
    + \sum_{i,j=1,2} \lambda_{H_{i}\phi_{j}} 
    (H^{\dagger}_{i} H_{i}) (\phi^{\dagger}_{j} \phi_j)
\label{potential}
\end{eqnarray} 

The scalars $H_{1}$, $H_{2}$, $\phi_{1}$ and $\phi_{2}$ take the
following form at the time  of $U(1)_{B_{i} - L_{i}}$, PQ and
electroweak symmetry, $SU(2) \times U(1)_{Y}$, breaking,
\begin{eqnarray}
&& H_{1}=
\begin{pmatrix}
H^{+}_{1} \\
\dfrac{v_1 +H^{0}_{1} + i A^{0}_{1}}{\sqrt{2}}
\end{pmatrix},
\,\,
H_{2}=
\begin{pmatrix}
H^{+}_{2} \\
\dfrac{v_2 +H^{0}_{2} + i A^{0}_{2}}{\sqrt{2}}
\end{pmatrix}
,\nonumber \\ && 
\phi_{1}=
\begin{pmatrix}
\dfrac{v_{\phi_1}+\phi^0_1}{\sqrt{2}} 
\end{pmatrix} e^{i \frac{a_1}{v_{\phi_1}}},
\phi_{2}=
\begin{pmatrix}
\dfrac{v_{\phi_2}+\phi^0_2}{\sqrt{2}} 
\end{pmatrix}
\label{phih}
\end{eqnarray}
The tadpole conditions which we obtain after demanding the first derivative
of the potential with respect to the neutral components of the 
scalar fields are zero,
\begin{eqnarray}
\mu^2_{11} &=& \frac{2 \lambda_{1} v^3_{1} + 
(\lambda_{12} + \lambda^{\prime}_{12}) v_{1} v^2_{2} 
+ \sqrt{2} \mu v_{2} v_{\phi_1} + v_{1} 
(\lambda_{H_{1}\phi_{1}} v^2_{\phi_1} + \lambda_{H_{1}\phi_{2}} v^2_{\phi_2})}{2 v_{1}}\,, \nonumber \\
\mu^2_{22} &=& \frac{2 \lambda_{2} v^3_{2} + 
(\lambda_{12} + \lambda^{\prime}_{12}) v_{2} v^2_{1} 
+ \sqrt{2} \mu v_{1} v_{\phi_1} + v_{2} 
(\lambda_{H_{2}\phi_{1}} v^2_{\phi_1} + \lambda_{H_{2}\phi_{2}} v^2_{\phi_2})}{2 v_{2}}\,, \nonumber \\
\mu^2_{\phi_1} &=& \frac{\lambda_{H_{1}\phi_{1}} v^{2}_{1} v_{\phi_1}
+ \lambda_{H_{2}\phi_1} v^2_{2} v_{\phi_1} + \sqrt{2} \mu v_{1} v_{2}
+ 2 \lambda_{\phi_1} v^3_{\phi_1} + \lambda_{\phi_1 \phi_2} v^2_{\phi_2}
v_{\phi_1} }{2 v_{\phi_1}}\,, \nonumber \\
\mu^2_{\phi_2} &=& \frac{\lambda_{H_{1}\phi_{2}} v^{2}_{1}
+ \lambda_{H_{2}\phi_2} v^2_{2} 
+ 2 \lambda_{\phi_2} v^2_{\phi_2} + \lambda_{\phi_1 \phi_2} v^2_{\phi_1}
}{2 }\,,
\end{eqnarray}

The neutral Higgs mass matrix in the basis 
($H^{0}_{1}\,\,\,H^{0}_{2}\,\,\,\phi^0_{2}\,\,\,\phi^0_{1}$) takes the
following form,
\begin{eqnarray}
\mathcal{L}^{NH} &=& 
\begin{pmatrix}
H^{0}_{1} & H^{0}_{2} & \phi^0_{2} & \phi^0_{1}
\end{pmatrix}
M^2_{S}
\begin{pmatrix}
H^{0}_{1} \\
H^{0}_{2} \\
\phi^0_{2} \\
\phi^0_{1}
\end{pmatrix} \,,
\end{eqnarray} 
where the neutral Higgs mass matrix, $M^2_{S}$, takes the following form,
\begin{eqnarray}
M^2_{S} = 
\begin{pmatrix}
2 \lambda_{1} v^2_{1} - \frac{\mu v_{2} v_{\phi_1}}{\sqrt{2} v_{1}} &
 \left( (\lambda_{12} + \lambda^{\prime}_{12})v_{1} v_{2} 
+ \frac{\mu v_{\phi_1}}{\sqrt{2}} \right) & {\lambda_{H_{1}\phi_2} v_{1} v_{\phi_2}}
 &  \left( \frac{\mu v_{2}}{\sqrt{2}} + \lambda_{H_{1} \phi_1}
v_{1} v_{\phi_1}\right)  
\\
  \left( (\lambda_{12} + \lambda^{\prime}_{12})v_{1} v_{2} 
+ \frac{\mu v_{\phi_1}}{\sqrt{2}} \right)
&
\left(2 \lambda_{2} v^2_{2} - \frac{\mu v_{1} v_{\phi_1}}{\sqrt{2} v_2} \right) &
{\lambda_{H_{2}\phi_2} v_{2} v_{\phi_2}}
 & \left( \frac{\mu v_{1}}{\sqrt{2}} + \lambda_{H_{2}\phi_1} v_{2}
v_{\phi_1} \right)
 \\
{\lambda_{H_{1}\phi_2} v_{1} v_{\phi_2}} & {\lambda_{H_{2}\phi_2} v_{2} v_{\phi_2}} &
2 \lambda_{\phi_2} v^2_{\phi_2} & {\lambda_{\phi_1 \phi_2} v_{\phi_1} v_{\phi_2}} \\
 \left( \frac{\mu v_{2}}{\sqrt{2}} + \lambda_{H_{1} \phi_1}
v_{1} v_{\phi_1}\right) & 
  \left( \frac{\mu v_{1}}{\sqrt{2}} + \lambda_{H_{2}\phi_1} v_{2}
v_{\phi_1} \right) &
 {\lambda_{\phi_1 \phi_2} v_{\phi_1} v_{\phi_2}} & 
 -\frac{\mu v_{1} v_{2}}{\sqrt{2} v_{\phi_1}} + 2 \lambda_{\phi_1} v^2_{\phi_1}
 \end{pmatrix} \nonumber \\
 \label{CP-even-mass}
\end{eqnarray}
As will be discussed in Section \ref{axion-part} to have a negligible contribution to SM Higgs mass from the other heavy Higgses through higher 
loop order correction terms, we need to take 
$\lambda_{i\phi_{1}} \rightarrow 0$ ($i = H_{1},H_{2},\phi_{2}$) and $\mu \ll 1 $. 
This will make sure the $M^2_{S}$ 
mass matrix in the two block diagonal form as $(4\times4) \simeq (3\times3)\,\, \bigoplus\,\, (1\times1)$. Therefore, we can
relate the mass eigenbasis and flavour eigenbasis by the Unitary matrix as,
\begin{eqnarray}
\begin{pmatrix}
    h_{1}\\h_{2}\\h_{3}\\ \phi_{1}
        \end{pmatrix}
        = 
        \begin{pmatrix}
            R & 0 \\
             0& 1
        \end{pmatrix}
        \begin{pmatrix}
            H^{0}_{1} \\ H^{0}_{2} \\ \phi^{0}_{2} \\ \phi^{0}_{1}
        \end{pmatrix}
\end{eqnarray}
where 
\begin{eqnarray}
    R = \begin{pmatrix}
    c_{1}c_{2} & s_{1} c_{2} & s_{2} \\
            -c_{1} s_{2} s_{3} - s_{1} c_{3} & c_{1}c_{3} -s_{1}s_{2}s_{3} & c_{2}s_{3}  \\
            -c_{1} s_{2} c_{3} + s_{1} s_{3} & - c_{1} s_{3} -s_{1} s_{2} c_{3} & c_{2}c_{3}    
    \end{pmatrix}
    \label{pmns-matrix}
\end{eqnarray}
and $c_{i} = \cos\alpha_{i}$, $s_{i} = \sin\alpha_{i}$ and 
$-\frac{\pi}{2} \leq \alpha_{i} \leq \frac{\pi}{2}$ ($i = 1,2,3$). The above diagonalisation 
matrix can go back to the 
original 2HDM diagonalization matrix, with the singlet scalars completely decoupled from the two doublets,
once we choose
$\alpha_{1} \rightarrow \alpha $ and $\alpha_{2,3} \rightarrow 0$. 

The charged Higgs mass matrix in the basis $(H^{+}_{1}\,\,\,H^{+}_{2})$
takes the following form,
\begin{eqnarray}
M^2_{\pm} = 
\begin{pmatrix}
- \frac{v_{2}}{2 v_{1}} \left( \lambda^{\prime}_{12} v_{1} v_{2}
 + \sqrt{2} \mu v_{\phi_1} \right) & 
 \frac{\lambda^{\prime}_{12} v_{1} v_{2} + \sqrt{2} \mu v_{\phi_1}}{2} \\
 \frac{\lambda^{\prime}_{12} v_{1} v_{2} + \sqrt{2} \mu v_{\phi_1}}{2} &
 - \frac{v_{1}}{2 v_{2}} \left( \lambda^{\prime}_{12} v_{1} v_{2}
 + \sqrt{2} \mu v_{\phi_1} \right)\,.
\end{pmatrix}
\label{charge-higgs-mass}
\end{eqnarray}
Once we diagonalise Eq. (\ref{charge-higgs-mass}) then we have 
two mass eigenstates $(H^{\pm}\,\,G^{\pm})$ and they are related with the charged eigen basis as
\begin{eqnarray}
\begin{pmatrix}
    G^{\pm}\\H^{\pm}
\end{pmatrix}
=
\begin{pmatrix}
    \cos\beta & \sin\beta \\
    -\sin\beta & \cos\beta
\end{pmatrix}
\begin{pmatrix}
    H^{\pm}_{1}\\H^{\pm}_{2}
\end{pmatrix}
\end{eqnarray}
where $\tan\beta = \frac{v_{2}}{v_{1}}$. The mass of $H^{\pm}$ and $G^{\pm}$ 
in terms of the vevs take 
the following form,
\begin{eqnarray}
M^2_{H^{\pm}} = - \frac{v^2_{1} + v^2_{2}}{2} 
\left( 
\lambda^{\prime}_{12} + \frac{\sqrt{2} \mu v_{\phi_1}}{v_{1} v_{2}} 
\right)\,,\,\,\, M^2_{G^{\pm}} = 0
\end{eqnarray} 
Additionally, among the charged eigenstates, $G^{\pm}$ is the Goldstone boson 
which acts as the longitudinal mode of the $W^{\pm}$ gauge boson. 

The CP-odd Higgs mass matrix in the basis $(A_{1}\,\,A_{2}\,\,a_1)$
takes the following form,
\begin{eqnarray}
M^2_{A} = 
\begin{pmatrix}
- \frac{\mu v_{2} v_{\phi_1}}{\sqrt{2} v_{1}} & \frac{\mu v_{\phi_1}}{\sqrt{2}} & \frac{\mu v_{2}}{\sqrt{2}} \\
\frac{\mu v_{\phi_1}}{\sqrt{2}} & - \frac{\mu v_{1} v_{\phi_1}}{\sqrt{2} v_{2}} & - \frac{\mu v_{1}}{\sqrt{2}} \\
\frac{\mu v_{2}}{\sqrt{2}} & - \frac{\mu v_{1}}{\sqrt{2}} & 
- \frac{\mu v_{1} v_{2}}{\sqrt{2} v_{\phi_1}}
\end{pmatrix}
\label{cp-odd-matrix}
\end{eqnarray}
The other CP-odd component $a_2$ becomes the longitudinal component of the additional gauge boson 
coming from $U(1)_{B_{i}-L_{i}}$ gauge symmetry and does not contribute in the 
CP-odd scalar mass matrix, $M^2_{A}$. After diagonalising the mass matrix shown 
by Eq. (\ref{cp-odd-matrix}), 
we get the following masses,
\begin{eqnarray}
M^2_{A} = - \frac{\mu v_{\phi_1}}{\sqrt{2}} 
\left( \frac{v_1}{v_2} + \frac{v_2}{v_1} + \frac{v_1 v_2}{v^2_{\phi_1}} \right)\,, M^2_{G^0} = 0\,,\,\, M^2_{a} = 0\,
\end{eqnarray}
The mass eigenbasis ($A\,\,G^{0}\,\,a$) and flavour eigenbasis ($A_{1}\,\,A_{2}\,\,a_{1}$) can be related by the 
following matrix relation,
\begin{eqnarray}
    \begin{pmatrix}
        G^{0}\\A\\a
    \end{pmatrix}
    = U
    \begin{pmatrix}
        A_{1} \\ A_{2} \\ a_{1}
    \end{pmatrix} 
 \label{cp-odd-diagonaliser}   
\end{eqnarray}
where
\begin{eqnarray}
U
    = \frac{1}{\sqrt{\sec^{2}\beta + \left( \frac{v_{2}}{v_{\phi_{1}}}\right)^{2} }}
    \begin{pmatrix}
     \cos\beta \sqrt{\sec^{2}\beta + \left( \frac{v_{2}}{v_{\phi_{1}}}\right)^{2} } & 
        \sin\beta \sqrt{\sec^{2}\beta + \left( \frac{v_{2}}{v_{\phi_{1}}}\right)^{2} } & 0 \\
        -\tan\beta & 1 & \frac{v_{2}}{v_{\phi_{1}}} \\
        -\frac{v_{2}}{v_{\phi_{1}}} \sin\beta & \frac{v_{2}}{v_{\phi_{1}}} \cos\beta & -\sec\beta 
    \end{pmatrix}
\end{eqnarray}

The CP-odd state $A$ has the mass and among the remaining two, $G^0$
is the Goldstone boson associated with the SM $Z$-boson and $a$ is the 
axion field which will get mass during the QCD phase transition \cite{DiLuzio:2020wdo}. 

The 2HDM with a singlet scalar extension in the context of collider constraints has been studied in
\cite{Muhlleitner:2016mzt, Arhrib:2018qmw}. This work focuses on the other aspects of the 
2HDM extension, namely dark matter, 
$(g-2)_{\mu}$, neutrino mass and excess contribution in W-boson mass after adding two singlet
scalars and an abelian gauge symmetry. The parameters have been chosen in such a way that
they comply with current collider constraints but will be in reach of future 
increased experimental sensitivity. 
The coupling of the Higgses with the SM gauge bosons can be expressed as,
\begin{eqnarray}
    \mathcal{L}_{h_{i}VV}  = i g_{\mu\nu} \lambda_{h_{i}VV} g^{SM}_{VV}
\end{eqnarray}
where $V = W^{\pm}, Z$, $\lambda_{h_{i}VV} = \cos\beta R_{i1} + \sin\beta R_{i2}$ and $g^{SM}_{VV}$
is the SM Higgs coupling with the gauge bosons in pure SM. The coupling of three Higgses 
with the gauge boson can be summarised as,
\begin{eqnarray}
    g_{h_{1}VV} &=& c_{\alpha_{2}} c_{\beta - \alpha_{1}} \nonumber \\
        g_{h_{2}VV} &=& c_{\alpha_{3}} s_{\beta - \alpha_{1}} - s_{\alpha_{2}} s_{\alpha_{3}} c_{\beta - \alpha_{1}}  \nonumber \\
            g_{h_{3}VV} &=& -s_{\alpha_{3}} s_{\beta - \alpha_{1}} - s_{\alpha_{2}} c_{\alpha_{3}} c_{\beta - \alpha_{1}}.
\end{eqnarray}
In the limit $\beta -\alpha_{1} = \frac{\pi}{2}$, $\alpha_{2,3} \rightarrow 0$, we go back to 
the 2HDM misalignment limit and $h_{2}$ coincide with the SM Higgs. In our analysis, we 
are going to consider $h_{2}$ as the SM Higgs field and fix its mass at $M_{h_2} = 125.5\,\,GeV$.
We are going to summarise the scalars couplings to fermion which will be needed for studying the muon  $(g-2)$. Therefore, the Lagrangian associated with the scalars and the fermions
take the following form,
\begin{eqnarray}
    \mathcal{L}_{Yuk} &=& - \sum_{f} \left( \lambda^{h_{i}}_{u} \frac{m_{f}}{v} \bar{f} f h_{i} 
    - i \lambda^{A_j}_{f} \frac{m_{f}}{v} \Bar{f} \gamma^{5} f A_{j}\right) \nonumber \\
    &+& \left[ {\sqrt{2} V_{ud}} H^{+} \bar{u} 
    \left( y^{H^{\pm}}_{u} \frac{m_{u}}{v} P_{L} 
    + y^{H^{\pm}}_{d} \frac{m_{d}}{v} P_{R}\right)d + y^{H^{\pm}}_{l} 
    \frac{\sqrt{2} m_{l}}{v} H^{+} \bar{\nu}_{L} l_{R}  + {\it h.c.} \right]
    \label{Higgs-fermion-coefficients}
\end{eqnarray}
where the coefficients $y^{\phi}_{f}$ are shown in Table \ref{tab-yfphi-1} and they become equal
to $1$ in the misalignment limit described before for the SM Higgs field $h_2$.

\begin{table}[]
\begin{center}
\begin{tabular}{||c|  c| c| c| c| c| c| c| c| c||} 
 \hline
 Type & $\lambda^{h_{i}}_{u}$ & $\lambda^{h_{i}}_{d}$ & $\lambda^{h_{i}}_{l}$ 
 & $\lambda^{A_{j}}_{u}$ & $\lambda^{A_{j}}_{d}$ & $\lambda^{A_{j}}_{l}$ & $\lambda^{H^{\pm}}_{u}$ & $\lambda^{H^{\pm}}_{d}$ & $\lambda^{H^{\pm}}_{l}$ \\  
 \hline\hline
 I & $\frac{R_{i2}}{\sin\beta}$ & $\frac{R_{i2}}{\sin\beta}$ & $\frac{R_{i2}}{\sin\beta}$ & $\frac{U_{i2}}{\sin\beta}$ & $-\frac{U_{i2}}{\sin\beta}$ & $-\frac{U_{i2}}{\sin\beta}$ & $\cot\beta$ & $-\cot\beta$ & $-\cot\beta$ \\ 
 \hline
 II & $\frac{R_{i2}}{\sin\beta}$ & $\frac{R_{i1}}{\cos\beta}$ & $\frac{R_{i1}}{\cos\beta}$
 & $\frac{U_{i2}}{\sin\beta}$ & $-\frac{U_{i1}}{\cos\beta}$ & $-\frac{U_{i1}}{\cos\beta}$ & $\cot\beta$ & $\tan\beta$ & $\tan\beta$\\
 \hline
 X & $\frac{R_{i2}}{\sin\beta}$ & $\frac{R_{i2}}{\sin\beta}$ & $\frac{R_{i1}}{\cos\beta}$
 & $\frac{U_{i2}}{\sin\beta}$ & $-\frac{U_{i2}}{\sin\beta}$ & $-\frac{U_{i1}}{\cos\beta}$
 & $\cot\beta$ & $-\cot\beta$ & $\tan\beta$\\
 \hline
 Y & $\frac{R_{i2}}{\sin\beta}$ & $\frac{R_{i1}}{\cos\beta}$ & $\frac{R_{i2}}{\sin\beta}$
 & $\frac{U_{i2}}{\sin\beta}$ & $-\frac{U_{i1}}{\cos\beta}$ & $-\frac{U_{i2}}{\sin\beta}$
 & $\cot\beta$ & $\tan\beta$ & $-\cot\beta$\\ 
 \hline
\end{tabular}
\caption{Coeffients of the scalars with the fermions where the full Lagrangian is represented 
in Eq. \ref{Higgs-fermion-coefficients}. In the table, the indices $i,j$ for 
the CP-even and CP-odd scalars vary as $i=1,2,3$ and $j=2,3$.}
\label{tab-yfphi-1}
\end{center}
\end{table}

\section{Neutrino mass}
\label{NM-Type-I}
After the symmetry breaking, the Lagrangian as shown in Eq.\,(\ref{lagN})
gives us the neutrino mass matrix in the basis 
$\left(\nu^c_L\,\,\,\, N_{f}\right)$ as follows,
\begin{eqnarray}
\mathcal{L}_{neutrino} &=& 
\begin{pmatrix}
\bar \nu_{L} & \bar N^c_{f}
\end{pmatrix}
\begin{pmatrix}
0 & M^T_{D} \\
M_{D} & M_{R}
\end{pmatrix}
\begin{pmatrix}
\nu^c_L\\
N_f
\end{pmatrix}
+ {\it h.c.}
\label{neutrino-mass}
\end{eqnarray}
where the Dirac mass matrix ($M_{D}$) and Majorana mass matrix 
($M_{R}$) takes the following form,
\begin{eqnarray}
M_{D}  = 
\begin{pmatrix}
\frac{y_{ee} v}{\sqrt{2}} & 0 & 0 \\
0 & \frac{y_{\mu\mu} v}{\sqrt{2}} & 0 \\
0 & 0 & \frac{y_{\tau\tau} v}{\sqrt{2}} \\
\end{pmatrix}\,,
\,\,\,\,\,
M_{R} = \frac{v_{\phi_1}}{\sqrt{2} M_{pl}}
\begin{pmatrix}
M_{ee} & \frac{\lambda_{e\mu} v_{\phi_2}}{\sqrt{2}} & M_{e\tau} \\
\frac{\lambda_{e\mu} v_{\phi_2}}{\sqrt{2}} & 0 & 
\frac{\lambda_{\mu\tau} v_{\phi_2} e^{i\theta}}{\sqrt{2}} \\
M_{e\tau} & \frac{\lambda_{\mu\tau} v_{\phi_2} e^{i\theta}}{\sqrt{2}} & 0
\end{pmatrix}\,.
\label{direc-majorana-mass-matrices}
\end{eqnarray}
Once we diagonalise Eq.\,(\ref{neutrino-mass}) in the seesaw limit
{\it i.e.} $M_{D} \ll M_{R}$, then we get light and heavy 
neutrino mass matrices as follows,
\begin{eqnarray}
m_{\nu} = - M^T_{D} M^{-1}_{R} M_{D}\,,\,\,\,M_{N} = M_{R}\,. 
\end{eqnarray}
Furthermore, upon the diagonalisation of the $m_{\nu}$ we get
three masses for the active neutrinos ($m^i_{d}$, i = 1,2,3) 
and diagonalisation matrix referred as PMNS matrix \cite{Maki:1962mu}, $U_{PMNS}(\theta_{12},\theta_{13},\theta_{23})$, which are related as follows,
\begin{eqnarray}
m_{d} = U^{T}_{PMNS} (\theta_{12},\theta_{13},\theta_{23})
\,m_{\nu}\, U_{PMNS} (\theta_{12},\theta_{13},\theta_{23})\,.
\end{eqnarray} 
where $\theta_{ij}\,(i,j = 1,2,3)$ are the oscillation angles.
Finally, the matrix $M_{N}$ represents the heavy neutrino mass matrix. 
It is worth pointing out that due to the additional
abelian gauge symmetry, we have just enough free parameters to accommodate values for the oscillation parameters in the correct ballpark of values favoured by the different oscillation experiments. To achieve such result, however, all the three right-handed neutrinos should participate to the see-saw mechanism. In other words, we cannot accommodate, in our setup, sterile neutrino DM by decoupling one of the right-handed neutrinos. This is an additional motivation, besides the interest on its own, for the introduction of axion DM.
 We also remark that the PQ controls the configuration of the Yukawa couplings of the Higgs doublets to the SM fermions.
In the present work, we have 
considered lepton-specific (Type-X) and Type-II 2HDM. 
Let's finally move to neutrino masses.
If we consider Universal Yukawa couplings for the Dirac and Majorana
mass terms then we can write down them as,
\begin{eqnarray}
    M_{D} = \frac{y_{ii} v_{2}}{\sqrt{2}} \,\,(i=e, \mu, \tau)\,\,\,{\rm and}\,\,\,
    M_{N} = \frac{\lambda_{ij} v_{\phi_{2}} v_{\phi_{1}} }{2 M_{pl}}
    = \frac{\lambda_{ij} M_{Z^{\prime}} v_{\phi_{1}} }{2 g^{\prime} M_{pl}}\,\,(i,j = e, \mu, \tau) 
    \,. 
    \label{Dirac-Majorana-mass-term}
\end{eqnarray}
Therefore, the light neutrino mass matrix will take the form,
\begin{eqnarray}
    m_{\nu} &=& \frac{y^{2}_{ii} v^{2}}{2 M_{N}} \left( \frac{\tan^{2}\beta}{1 + \tan^{2}\beta} \right)
    \nonumber \\
    &=& \frac{y^2_{ii} v^{2} g^{\prime} M_{pl}}{\lambda_{ij} M_{Z^{\prime}} v_{\phi_{1}} }
    \left( \frac{\tan^{2}\beta}{1 + \tan^{2}\beta } \right)\,.
    \label{neutrino-mass-dependence}
\end{eqnarray}
Neutrino data from different measurements have put severe constraints on the mixing angles of
the PMNS matrix and the mass square differences. Current bound on the mixing angles for Normal
ordering in $3\sigma$ range is give by \cite{Esteban:2020cvm},
\begin{eqnarray}
    \theta^{0}_{12} \rightarrow \left( 31.31 \rightarrow 35.74  \right)\,,
    \theta^{0}_{13} \rightarrow \left( 8.19 \rightarrow 8.89  \right)\,,
    \theta^{0}_{23} \rightarrow \left( 39.60 \rightarrow 51.90  \right)\,,
\end{eqnarray}
and the mass square differences as,
\begin{eqnarray}
    \Delta m^2_{21} \rightarrow \left( 6.82 \rightarrow 8.03 \right) \times 10^{-5} \,\,{\rm eV^{2}}\,,
        \Delta m^2_{31} \rightarrow \left( 2.428 \rightarrow 2.597 \right) \times 10^{-3} \,\,{\rm eV^{2}}\,.
\end{eqnarray}
The ranges for the inverted ordering are also similar but $\Delta m^{2}_{31}$ is with a negative
sign. For the present structure of the neutrino mass ordering, one of the authors
has already studied the neutrino mass and obtained the correct values of the oscillation 
parameters and showed that inverted mass ordering is not possible for this kind of mass matrix
structure \cite{Biswas:2016yan}. In the present work, we assume that the correct ranges 
of oscillation parameters can be achieved and instead study the range of the parameters
which will give the feasible values for the Dirac and Majorana mass matrix namely if 
the associated Yukawa coupling for the Dirac mass term can be made larger than the electron
Yuwaka coupling and the Majorana mass terms above the keV range. This is necessary to
check because the Planck suppresses some of the elements in the Majorana mass matrix 
mass term. 

\begin{figure}[h!]
\centering
\includegraphics[angle=0,height=7.5cm,width=7.5cm]{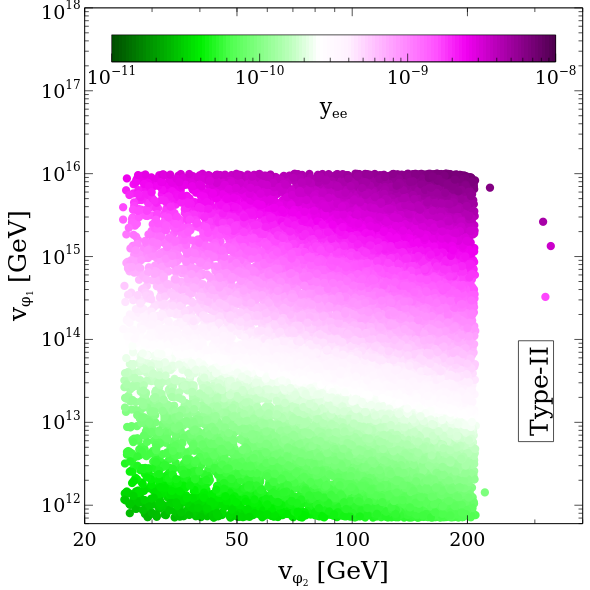}
\includegraphics[angle=0,height=7.5cm,width=7.5cm]{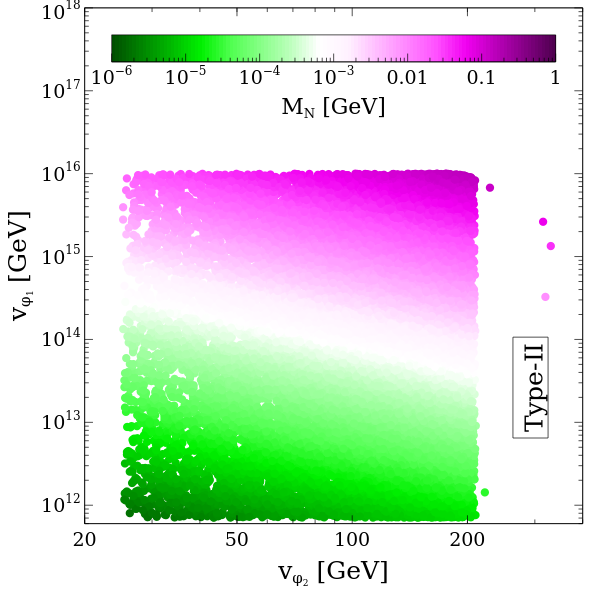} \\
\includegraphics[angle=0,height=7.5cm,width=7.5cm]{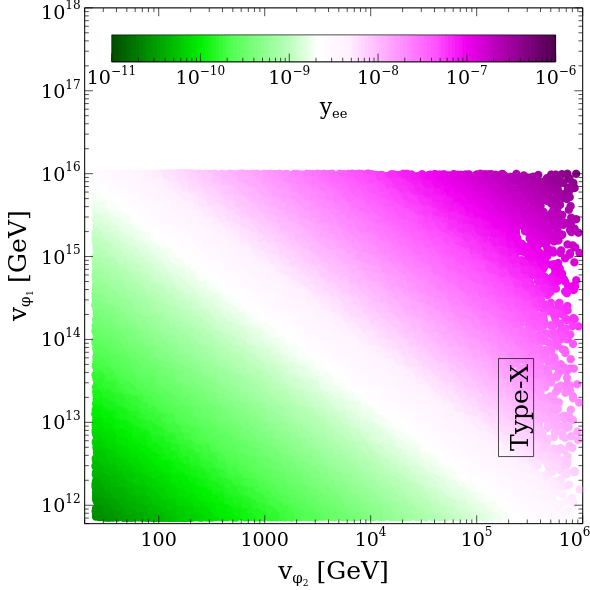}
\includegraphics[angle=0,height=7.5cm,width=7.5cm]{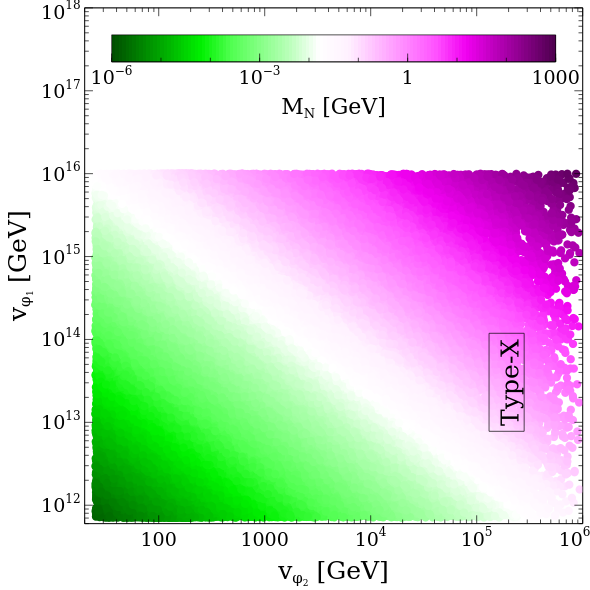}
\caption{Scatter plots in $v_{\phi_{2}}-v_{\phi_{1}}$ plane where in the LP colour variation corresponds 
to the Dirac Yukawa coupling for the neutrinos and RP shows the colour variation for the 
right-handed neutrino mass. The upper panel and lower panel correspond to the Type-II and Type-X 2HDM models.} 
\label{neutrino-plots}
\end{figure}

In the left figure and right figure of the upper panel, we have shown scatter plots in the 
$v_{\phi_{2}}-v_{\phi_{1}}$ plane. The colour variation in the left one is for the Dirac Yukawa coupling
associated with the neutrinos and the right figure is for the right-handed neutrino mass. Both
the plots are for the Type-II 2HDM model. The lower panel represents the same plots but for the 
Type-X 2HDM model. In generating the plots, we have considered the order of the neutrino mass
$m_{\nu} = 10^{-11}$ GeV (as shown in Eq. \ref{neutrino-mass-dependence}) and the Yukawa 
couplings for the right-handed neutrino mass matrix (shown by Eq. \ref{Dirac-Majorana-mass-term}) 
as $\lambda_{e\mu} = \lambda_{\mu\tau} = 1$. 
This is worth mentioning that all the points in the plots can give us
the correct value of muon $g-2$, the total amount of DM by suitably choosing the initial angle 
and also contributing to the oblique parameters $S, T, U$ which will be addressed in the next sections.
In the left figure of the upper panel, we can see that 
$v_{\phi_2}$ can not be larger than 210 GeV and $v_{\phi_1}$ can take the values up to maximal
varied range $10^{16}$ GeV. In the figure, we can see that for a fixed value $v_{\phi_1}$ if
we increase the $v_{\phi_2}$ values then the Yukawa coupling $y_{ee}$ starts increasing
as depicted by the colour variation. The variation in $y_{ee}$ is not huge because $v_{\phi_2}$ does not 
vary in a large range for the Type-II 2HDM case. On the other hand, if we move towards the 
increment of $v_{\phi_1}$ direction
then we observe a large variation in $y_{ee}$ because of the large varied range in $v_{\phi_1}$.
In the right figure of the upper panel, we can see the colour variation in right-handed neutrino mass.
The right-handed neutrino masses depends on the vevs $v_{\phi_{1,2}}$ as 
$M_{N} \propto v_{\phi_{1}} v_{\phi_2} $, therefore we see a anticorrelation between $v_{\phi_1}$
and $v_{\phi_2}$ if we take a fixed value of right-handed neutrino mass. Moreover, we see increment 
in $M_{N}$ as we move in any incremental direction of the vevs $v_{\phi_{1,2}}$. From both
the figure, we can see that the right-handed neutrino mass can go to a maximum 
value of 1 GeV and the associated Dirac
Yukawa coupling reaches at most $10^{-8}$. Therefore, for the Type-II case,
we can not achieve $y_{ee}$ up to
the value of the Yukawa coupling needed for the electron mass in SM. On the other hand,
in the lower panel, we have shown the same thing but for large values of $v_{\phi_2}$ as well, this helps us in
getting the $y_{ee}$ as large as electron Yukawa coupling and the right-handed neutrino mass
as large as 1 TeV. As we will see later for the Type-X 2HDM, muon $g-2$ can also achieved only by 
the Higgs sector which helps us to lower the gauge coupling $g^{\prime}$
and hence obtained the increased value of the vev $v_{\phi_2} = \frac{M_{Z^{\prime}}}{g^{\prime}}$.
Finally, low-mass right-handed neutrinos will be constrained by many 
experiments and will also be explored in future by a number of experiments. 
A detailed discussion of the 
low mass right-handed neutrino detection prospects can be found in \cite{Deppisch:2015qwa}. 
Moreover, as shown in Ref. \cite{Deppisch:2015qwa}
right-handed neutrino mass below 0.7 GeV will impact the BBN bound 
\cite{Gorbunov:2007ak, Boyarsky:2009ix, Ruchayskiy:2012si}. This problem can be 
overthrown in the present context by choosing the higher values of $M_{ee}, M_{\mu\tau}$ 
(compatible with the neutrino oscillation data as shown in \cite{Biswas:2016yan})
parameters in the right-handed neutrino
mass matrix (see Eq. \ref{direc-majorana-mass-matrices}) which makes the right-handed 
neutrino masses higher and can evade the BBN bound.  
For precise statement in this context, we require full-fledged study and 
left to pursue in future. The baryogenesis process concerning the MeV scale right-handed neutrino mass within the Type-I seesaw mechanism has been studied in Ref. \cite{Domcke:2020ety} albeit the right-handed neutrino decay needed before BBN.

\section{Axion DM}
\label{axion-part}

In this work, we consider that the singlet scalar $\phi_1$ takes spontaneous vev ($v_{\phi_1}$) which is around or above the inflation scale. 
This high vev ensures that our axion 
field would be invisible to the visible sector due to its suppressed coupling 
impacted by the very high vev. This also exhibits that the associated 
neutral CP-even Higgs would also be very heavy and can be approximated as
$M^2_{\phi_1} \simeq \lambda_{\phi_1} v^2_{\phi_1}$. This mass eigenstate 
has the mixing term with the SM Higgs and will contribute to Higgs mass 
radiatively as,
\begin{eqnarray}
M^2_{h_{2}} = M^{2}_{h_{2},tree} + \lambda_{H_{i}\phi_{1}} \left[ \Lambda^2_{UV}
+ M^2_{\phi_1} \log \left( \frac{M^2_{\phi_1}}{\Lambda^2_{UV}} \right) \right]\,.
\end{eqnarray}  
Here $M^2_{\phi_1}$ is a physical parameter and can not be removed by
introducing the counter terms. As discussed in \cite{Foot:2013hna}, 
if we go in the limit $\lambda_{H_{i}\phi_{1}} \rightarrow 0$ ($i = 1, 2$), then we go in the regime of the enhanced 
Poincar\'{e} symmetry and we can protect the light SM Higgs mass. The
bound on the quartic couplings which relate the SM Higs and $\phi_1$
is following,
\begin{eqnarray}
\lambda_{H_{i}\phi_{1}} \leq \mathcal{O}\left( \frac{M_{h}}{M_{\phi_{1}}} \right),\,\, (i=1, 2).
\end{eqnarray} 
Therefore, in this work, we consider $\lambda_{H_{i}\phi_{1}} \rightarrow 0$
and the field  $\phi_1$ couple to the SM sector very feebly and 
do not alter our phenomenology presented in this work.
Therefore, the neutral Higgs mass matrix as given in Eq. (\ref{CP-even-mass}) 
effectively reduces to $3\times 3$ matrices which can be diagonalised
by the PMNS matrix matrix $R$ shown in Eq. \ref{pmns-matrix}. 

\subsection{ PQ Symmetry:}

As discussed before in the present work, we can introduce the $U(1)_{PQ}$
symmetry with the charge assignment shown in Table \ref{tab3} and when it breaks
spontaneously we can have a massless axion field.
As pointed out in 
Refs. \cite{Hawking:1987mz, Lavrelashvili:1987jg, Giddings:1988cx, Coleman:1988tj, Gilbert:1989nq}, the global abelian symmetry is not a 
good symmetry at the Planck scale, so it is unpleasing to introduce 
a global $U(1)$ symmetry barely in the SM because it will not be valid at the
Planck scale. 
This situation can be evaded if PQ symmetry appears accidentally and can be
achieved by introducing the discrete symmetries as studied 
in \cite{Bjorkeroth:2017tsz, Harigaya:2013vja, Dias:2002hz, Ahn:2014gva, Honecker:2015ela}. 
Therefore, we can always 
decide on the discrete symmetries which ensures the accidental appearance of the PQ symmetry
as well as does not alter our phenomenology.
This way we can solve the potential problem regarding the validity of global PQ 
symmetry at the Planck scale.
In this work, the Higgs doublets and fermions are charged under the 
$U(1)_{PQ}$ symmetry which makes the present model
like DFSZ kind \cite{Zhitnitsky:1980tq, Dine:1981rt} axion model in contrary to the KSVZ type model
\cite{Kim:1979if, Shifman:1979if, Covi:2022hqb}
where one needs to introduce extra exotic quarks in order
to generate axion-gluon coupling. In the present work, we consider singlet-dominated
pseudoscalar ``$a$" as the axion field with axion decay constant $v_{a} = \sqrt{v^2_{\phi_1}
+ \frac{v^2_{1} v^2_{2}}{v^2}} \simeq v_{\phi_1}$, for $v_{\phi_1} \gg v, v_{1,2}$.
By using the diagonalisation matrix as shown in Eq.\,\ref{cp-odd-diagonaliser}, we can write down the $A_{1,2}$
in terms of the axion field like $\frac{A_{1}}{v_{1}} \rightarrow X_{H_1}\frac{a}{v_{a}}$,
$\frac{A_{2}}{v_{2}} \rightarrow X_{H_2}\frac{a}{v_{a}}$
and $\frac{a_{1}}{v_{\phi_1}} \rightarrow X_{\phi_1}\frac{a}{v_{a}}$ where $X_{F}$ is the PQ
charge of the field $F$ ($= H_{1,2}, \phi_{1}$). We can write down the mass term for the 
quarks and leptons consist of the axion field as,
\begin{eqnarray}
    \mathcal{L}_{a-mass} \supset - m_{u} \bar{u}_{L} u_{R} e^{- i X_{H_2} \frac{a}{v_{a}}}
    - m_{d} \bar{d}_{L} d_{R} e^{- i X_{H_m} \frac{a}{v_{a}}}
    - m_{l} \bar{l}_{L} l_{R} e^{- i X_{H_n} \frac{a}{v_{a}}} + {\it h.c.}\,.
\end{eqnarray}
We can now get rid of the phase factor terms consisting of the axion field by redefining the 
fields as,
\begin{eqnarray}
    u \rightarrow e^{i \gamma_{5} X_{H_{2}} \frac{a}{2 v_{a}}}\,u, \,\,\,
    d \rightarrow e^{-i \gamma_{5} X_{H_{m}} \frac{a}{2 v_{a}}}\,d,\,\,\,
    l \rightarrow e^{-i \gamma_{5} X_{H_{n}} \frac{a}{2 v_{a}}}\,l\,.
\end{eqnarray}
This way of redefining the fields will induce the anomaly associated with the QCD and 
electromagnetic (EM) fields. These anomaly terms can be written in terms of the 
QCD and EM fields as,
\begin{eqnarray}
    \mathcal{L}_{anomaly} = \frac{\alpha_{s}}{8 \pi} \frac{a}{f_{a}} G \Tilde{G}
    + \frac{\alpha_{EM}}{8 \pi} \frac{E}{N} \frac{a}{f_{a}} F \Tilde{F}
\end{eqnarray}
where $f_{a} = \frac{v_{a}}{|2 N|}$\,. $N$ and $E$ are respectively the anomaly 
factors depend on the QCD and EM charges coming from the 
$SU(3)_{c} \times SU(3)_{c} \times U(1)_{PQ}$ and $U(1)_{EM} \times U(1)_{EM} \times U(1)_{PQ}$
anomaly terms. The EM sector anomaly term $E$ and strong sector anomaly term $N$ can be 
expressed in terms of the PQ charges as \cite{DiLuzio:2020wdo},
\begin{eqnarray}
    N &=& 3 \left( -\frac{1}{2} X_{H_{2}} + \frac{1}{2} X_{H_{m}} \right)\,,\nonumber \\
    E &=& 3 \left( - 3 \left( \frac{2}{3} \right)^{2} X_{H_{2}} + 
    3 \left( - \frac{1}{3} \right)^{2} X_{H_{m}} + (-1)^{2} X_{H_{n}}\right)\,.
\end{eqnarray}
The values of $N$ and $E$ will depend on the $m,n$ values which represents the different kind of
2HDM as shown in Table \ref{tab5}.
\begin{table}[]
\begin{center}
\begin{tabular}{||c|  c| c | c ||} 
 \hline
 Type & $N$ & $E$ & E/N \\  
 \hline\hline
 I & $0$ & $0$ & $-$ \\  
 \hline
 II & $-3/2$ & $-4$ & $8/3$  \\
 \hline
 X & $0$ & $-3$ & $-$  \\
 \hline
Y & $-3/2$ & $-1$ & $2/3$  \\
 \hline
\end{tabular}
\caption{$N$ and $E$ values for different types of 2HDM models.}
\label{tab5}
\end{center}
\end{table}

The PQ charge of the SM and the BSM particles are given in Table \ref{tab3}.
As can be seen from Eq. \ref{potential}, the PQ symmetry 
breaks when the field $\phi_1$ gets vev and the axion potential is generated
at the QCD phase transition by the instanton effects.
Therefore, the periodic potential for the dynamical field axion can be expressed as \cite{DiLuzio:2020wdo},
\begin{eqnarray}
V_{QCD}\left( a \right) = 
M^2_{a} F^2_{a} \left( 1 - \cos\left[\theta_{a} + \bar{\theta} \right]  \right)
\label{axion-potential}
\end{eqnarray}       
where $F_{a}$ is axion decay constant, $\theta_{a} = \frac{a}{F_{a}}$ 
and $M_{a}$ is the axion mass given by \cite{DiLuzio:2020wdo},
\begin{eqnarray}
M^2_{a} &=& \frac{m_{u} m_{d}}{(m_{u} + m_{d})^{2}} \frac{f^2_{\pi} m^2_{\pi}}{F^2_{a}}\,\nonumber \\
&\simeq& 5.7\, \biggl( \frac{10^{12}\,\,{\rm GeV} }{F_{a}} \biggr)\,\,{\rm \mu\,eV} 
\end{eqnarray}
where $m_{u,d}$ are the quark masses, $f_{\pi} = 130$ MeV is the pion decay constant and
$m_{\pi} = 139.57$ MeV is the pion mass.
In the axion potential, we can define a new field after redefining it as $\bar{\theta_{a}} 
= \theta_{a} +\bar{\theta}$.
In Eq. \ref{axion-potential}, we have shown the axion potential
which will appear when the axion has coupling with the gluons, so in the present model
Type-II and Type-Y 2HDM fall into this category. Therefore, we can solve the strong CP
problem for these two variant of 2HDM model. Moreover, the axion appears from 
Type-II and Type-Y 2HDM can also be a good DM candidate which can be produced by 
the misalignment mechanism \cite{Turner:1985si}. 
On the other hand, Type-I and Type-X 2HDM models have no coupling
between axion and gluon fields so axion doesn't have potential as shown in Eq. \ref{axion-potential},
therefore they can not solve the strong CP problem. But they can be a good DM candidate and
produced by the misalignment mechanism \cite{Marsh:2015xka}. Moreover, their mass can be produced at the 
Planck scale by the PQ violating terms \cite{Covi:2022hqb} and we can produce them by the usual
misalignment mechanism. In this work, we have considered DM production only for the QCD-type
axion field and can also be estimated for the axion-like particle by following Ref. \cite{Marsh:2015xka}.

\subsection{ Axion density:}

The axion can be a very good cold DM candidate and is produced in the early
Universe by the misalignment mechanism \cite{Turner:1985si}. The amount of axion density 
depends on the axion decay constant ($F_{a}$) and the initial misalignment
angle $\theta_i$ and quantitatively can be expressed as \cite{Turner:1985si},
\begin{eqnarray}
\Omega_{a}h^{2} = 0.12\, \theta^2_{i}\, \left( \frac{F_{a}}{10^{12}\,\,{\rm GeV}} \right)^{1.19}
\label{misalignment-density}
\end{eqnarray} 
 If we consider the PQ symmetry breaking before the inflation scale
 then $\theta_i$ can take any arbitrary value and if it breaks
 after inflation, we need to take the average values of $\theta_i$ over the 
 different patches and takes the value \cite{GrillidiCortona:2015jxo},
 \begin{eqnarray}
 \theta_{i} \simeq \sqrt{\langle \theta^2_{i} \rangle} \sim 2.15\,.
\end{eqnarray}
In this work, we consider the PQ symmetry breaking before the inflation so 
we can ignore the contribution in the axion density from the 
cosmological defects otherwise one needs to take into account 
the axion density coming from the domain wall, strings \cite{Gorghetto:2018myk, Buschmann:2019icd,
Gorghetto:2020qws, Buschmann:2021sdq}.    
 
\subsection{ Isocurvature bounds:}

As studied in Refs. \cite{Kawasaki:2013ae, Kawasaki:2018qwp} if the PQ breaking happens during or before
the inflation then the massless axion can have quantum fluctuations 
which will contribute to the total energy density of the axion. 
The quantum 
fluctuation of the axion field is given by,
\begin{eqnarray}
\delta a = \frac{H_{inf}}{2\pi}
\end{eqnarray}  
where $\delta a = F_{a} \delta \theta_i$ and $H_{inf}$ is the Hubble parameter during inflation\,.

The axion fluctuation can generate the isocurvature perturbation, $S_{DM}$,
which is defined as \cite{Kawasaki:2013ae},
\begin{eqnarray}
S_{DM} = \Omega^{frac}_{a}h^{2}\, \frac{\delta\rho_{a}}{\rho_{a}}
\end{eqnarray} 
where $\Omega^{frac}h^{2} = \frac{\Omega_{a}h^{2}}{\Omega_{DM}h^{2}}$ is the 
fraction of the DM contained by the axion field. The CMB data \cite{Planck:2015sxf, Planck:2018jri}
has put a tight bound on the isocurvature perturbation which gives the bound in the inflation scale as \cite{Kawasaki:2018qwp},
\begin{eqnarray}
H_{inf} \leq 2.4 \times 10^{7} \,\,{\rm GeV}\,\,\left(\frac{F_{a}}{10^{12} \,\,{\rm GeV}} \right)^{0.405}\,.
\end{eqnarray}
The isocurvature perturbation also contributes to the axion density and changes
the Eq. \ref{misalignment-density} as follows,
\begin{eqnarray}
\Omega_{a}h^{2} = 0.18 \left[\theta^2_{i} + \left( \frac{H_{inf}}{2 \pi F_{a}}\right)^{2} \right] \left(\frac{F_{a}}{10^{12} \,\,{GeV}}  \right)^{1.18}\,.
\end{eqnarray} 

\begin{figure}[h!]
\centering
\includegraphics[angle=0,height=9.5cm,width=8.5cm]{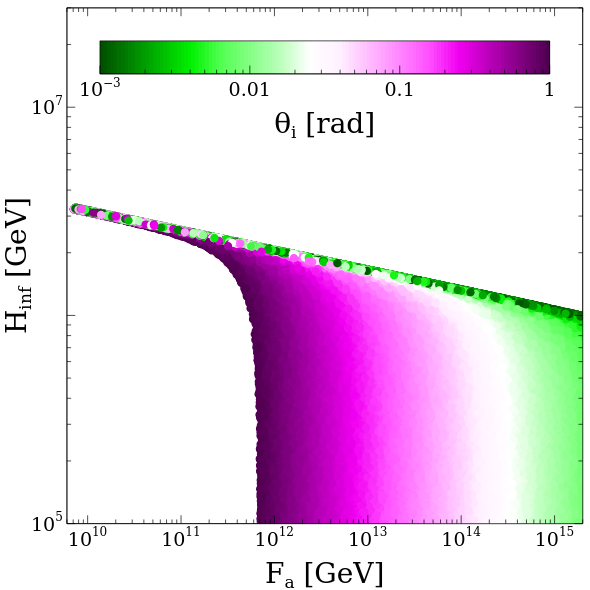}
\includegraphics[angle=0,height=9.5cm,width=8.5cm]{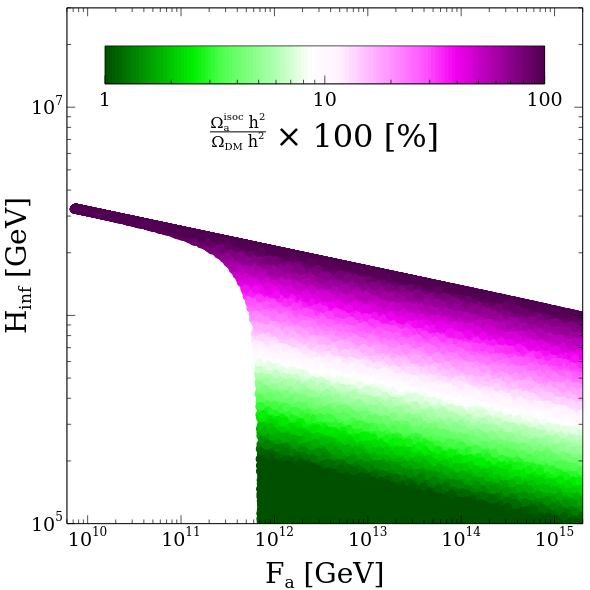}
\caption{Scatter plots in the $F_{A}-H_{inf}$ plane after satisfying the axion density in the 
$3\sigma$ range of dark matter relic density put by Planck. 
The colour bar corresponds to different values 
of $\theta_i$ and percentage of axion density from isocurvature fluctuation in the LP and RP,
respectively.} 
\label{HI-FA}
\end{figure}
In Fig. \ref{HI-FA}, we have shown scatter plots in the $F_{a}-H_{inf}$ plane where 
the colour variation in both panels is represented by $\theta_i$ and 
$\frac{\Omega^{isoc}_{a} h^{2}}{\Omega_{DM} h^{2}} \times 100\, [\%]$. In generating the plots,
we have varied the parameters in the following range,
\begin{eqnarray}
     10^{-3} \leq &\theta_{i}& \leq 1 \nonumber \\
    10^{5}\,\,{\rm GeV} \leq &H_{inf}& \leq 10^{7}\,\,{\rm GeV}
\end{eqnarray}
and the axion DM relic density has been demanded in the $3\sigma$ range put by Planck 
\cite{Planck:2015fie, Planck:2018vyg} {\it i.e.} 
$0.1172 \leq \Omega_{DM}h^{2} \leq 0.1226$.
In the LP, we look at the 
upper line begins at $H_{inf} = 3 \times 10^{6}$ GeV then it exhibits that the DM relic density is 
independent of $\theta_i$ for the range in which it has been varied. The line also implies that
if we take the inflation scale above this line, there will be an overproduction of DM. Below
this line or lower values of $H_{inf}$ we can see that DM production depends on $\theta_i$ and 
the axion produced mainly by the misalignment mechanism. We can see from the LP that if we increase 
the $F_{a}$ value then we need lower values of $\theta_i$ to satisfy the DM relic density
in the $3\sigma$ range. On the other hand in the RP, the colour bar corresponds to the contribution
to the axion density from the isocurvature perturbation. We can see if we take 
$H_{inf} > 3 \times 10^{6}$ GeV then the DM is overproduced from the isocurvature contribution and
as we go lower values of $H_{inf}$ then we get lower contribution in the axion density from axion 
fluctuation and more contribution starts coming from the misalignment mechanism.

\subsection{ Axion DM searches:}

There have been several attempts to detect the axion type DM at the haloscope experiments and also
proposals for future exploration in the context of the DFSZ and KSVZ types axion model.
A detailed discussion 
on the detection prospects by using different techniques has been 
exhibited in Ref. \cite{DiLuzio:2020wdo}.
In particular, Axion Dark Matter eXperiment (ADMX) has already probed the axion mass range 
$2.7-4.2$ $\mu eV$, using haloscope for axion dark matter, which represents the total amount of DM for $\mathcal{O}(1)$ misalignment 
angle \cite{ADMX:2018gho, ADMX:2019uok, ADMX:2021nhd}. There is also an attempt to explore the higher
mass range for example CAPP-8TB \cite{Lee:2019mfy} which aims to probe the axion mass range 
$6.62-7.04$ $\mu eV$ with the follow-up proposal like CAPP-12TB and CAPP-25T \cite{Semertzidis:2019gkj}.
Another dielectric haloscope experiment MADMAX \cite{MADMAX:2019pub} 
can probe the axion mass range $50-100$ $\mu eV$.
On the other hand, probing the lower mass range of axion DM needs a larger haloscope which is more
challenging. KLASH experiment \cite{Alesini:2019nzq} aims to explore 
the axion mass range $0.3-1$ $\mu eV$.
There is also an attempt to explore the axion mass range below $10^{-8}$ eV using different techniques.
ABRACADABRA \cite{Ouellet:2018beu} is one such attempt which can detect oscillating magnetic flux through the centre of the toroid produced by axion 
posed the possibility to detect the axion DM mass range below $10^{-8}$ eV. They have already
explored the mass range $(3.1-83.0) \times 10^{-10}$ eV from one-month data collection with
ABRACADABRA-10 cm.

\section{Muon ($g-2$)}
\label{muon-g-2}

In this section, we are going to address the muon ($g-2$) anomaly which SM can not explain 
but the present model has the potential to explain it. 
The SM contribution to $(g-2)_{\mu}$ has been measured very precisely after taking into 
account all the contributions namely quantum electrodynamics (QED), hadronic vacuum polarisation, hadronic
light by light and electroweak processes. Based on the aforementioned contributions, SM predicts the muon
($g-2$) contribution as 
$a_{\mu}$(SM) = $116591810(43)\times 10^{-11}$ (0.37 ppm) \cite{Aoyama:2020ynm}. 
Moreover, with the
advancement of experimental techniques, there is also an attempt to measure the muon ($g-2$) 
experimentally. The main contributors in the measurement of muon ($g-2$) experimentally include 
CERN \cite{Charpak:1962zz, Charpak:1965zz, Combley:1974tw, CERN-Mainz-Daresbury:1978ccd}, 
BNL E821 \cite{Muong-2:2006rrc} at Brookhaven and the recent measurement by the 
FNAL \cite{Muong-2:2021ojo}. BNL and FNAL
have followed the same techniques to measure the muon spin precision in the magnetic field
and agree to each other. The world average value in the present time from different experiments is 
$a_{\mu} (exp) = 116592061 (41)\times 10^{-11}$ (0.35 ppm). Therefore, the difference between the experimental and theoretical prediction from SM is \cite{Muong-2:2021ojo, Aoyama:2020ynm},
\begin{eqnarray}
    \Delta a_{\mu} &=& a_{\mu} (exp) - a_{\mu} (SM) \nonumber \\ 
&=& \left( 251 \pm 59  \right) \times 10^{-11}
\end{eqnarray}
which implies a $4.2\sigma$ discrepancy between the experimental and theoretical values. 
It is worth mentioning
here that the recent advancement in lattice computation 
measured the hadronic vacuum polarisation which differs from the different experimental measurements 
of the hadronic vacuum polarization by $2.1\,\,\sigma$ \cite{Borsanyi:2020mff}.
Therefore, it
reduces the significance of the
discrepancy between the experiment and theory to $1.5\sigma$ \cite{Borsanyi:2020mff}. Moreover, a new measurement by CMD-3 experiments for VHP contribution to the theoretical prediction indicates that the theoretical and experimental
values are closer than before with $2.4\sigma$ discrepancy \cite{CMD-3:2023alj}. 
This new finding also poses 
a conflict with the previous measurements of $e^{+}e^{-} \rightarrow \pi^{+}\pi^{-}$ by the same 
and different experiments 
\cite{Davier:2017zfy, Keshavarzi:2018mgv, Colangelo:2018mtw, Davier:2019can}. In summary, new data are required to definitively assess that the measure of $(g-2)$ really represents a deviation with respect to the SM prediction. Having this in mind we take, anyway, the hypothesis, for this work, that New Physics is responsible for the $g-2$ anomaly. 
In the scenario under scrutiny, there are two kinds of BSM contributions to the anomalous magnetic moment of the muon. The first come from the extended Higgs sector, see e.g. \cite{Arcadi:2022lpp, Arcadi:2022dmt, Arcadi:2021zdk, Arcadi:2021yyr}, while the other comes from the additional gauge boson \cite{Biswas:2016yan, Biswas:2016yjr, Biswas:2017ait, Biswas:2021dan, Costa:2022oaa}.
We will see that both the 
contributions are important in order to explain the muon ($g-2$). As we have explained before, 
we need a positive contribution from the BSM physics in order to address the $(g-2)_{\mu}$ anomaly.
In the case of scalar sector contribution, we have one loop and two loop contributions. The one loop contribution can be summarised as 
\cite{Lautrup:1971jf, Leveille:1977rc, Dedes:2001nx, Broggio:2014mna, Ilisie:2015tra},
\begin{eqnarray}
    \Delta a^{1-loop}_{\mu} = \frac{G_{F} m^2_{\mu}}{4 \pi^{2} \sqrt{2}} \sum_{j} (y^{j}_{\mu})^{2} 
    r^{j}_{\mu} f_{j} (r^{j}_{\mu})
\end{eqnarray}
where $j = \{h_{1},h_{2},h_{3},A,a,H^{\pm} \}$, $G_{F}$ is Fermi constant, $r^{j}_{\mu} = \frac{m^{2}_{\mu}}{M^2_{j}}$ ($m_{\mu}$ is muon mass and $M_{j}$ is the mediator scalar mass) 
and the 
function $f_{j}(r)$ has the following form,
\begin{eqnarray}
    f_{h_{1,2,3}}(r) &=& \int^{1}_{0} dx \frac{x^{2} (2-x)}{1-x+rx^{2}}\,, \nonumber \\
    f_{A,a}(r) &=& \int^{1}_{0} dx \frac{-x^{3}}{1-x+rx^{2}} \,, \nonumber \\
    f_{H^{\pm}}(r) &=& \int^{1}_{0} \frac{-x (1-x)}{1-(1-x)r}\,.
\end{eqnarray}
The normalised Yukawa coupling is defined as
$L_{jff} = y^{j}_{\mu} \left( \frac{m_{f}}{v} \right) j\,\bar{f} f$ and given in Table \ref{tab3}.
The contribution from the scalar sector comes from the diagrams mediated by $h_{2,3}, A, a$
and we have ignored the contribution from the SM like Higgs $h_{1}$ because we 
work in the misalignment limit $\beta - \alpha = \frac{\pi}{2}$ and its contribution is already 
taken in the standard computation of $\Delta a_{\mu}$.

From the previous expressions of the one-loop contribution to $\Delta a_{\mu}$, we can see that
the contribution is proportional to the fourth power of the muon mass. 
It has been shown in \cite{Chang:2000ii, Cheung:2001hz, 
Cheung:2003pw}
that the two-loop contribution which is proportional to the heavier fermions can contribute significantly
compared to the one-loop contribution. The two-loop contribution comes from the Bar-Zee type diagram 
and contributes to the $(g-2)_{\mu}$ by the following amount \cite{Chang:2000ii, Cheung:2001hz, 
Cheung:2003pw},
\begin{eqnarray}
    \Delta^{2-loop}_{\mu} = \frac{G_{F} m^2_{\mu}}{4 \pi^{2} \sqrt{2}} \frac{\alpha_{EM}}{\pi}
    \sum_{j,f} N^{c}_{f} Q^{2}_{f} y^{j}_{\mu} y^{j}_{f} r^{j}_{f} g_{i}(r^{j}_{f})
\end{eqnarray}
where $j$ is as defined earlier and $f = b$-quark, $t-$quark and $N^c_{f}, y^j_{f}, Q_{f}$
represent the colour degree of freedom, associated  Yukawa coupling of fermion $f$ with 
the scalar $j$, electromagnetic charge of mediator fermion $f$. The function $g_{i}(x)$ can be 
expressed as,
\begin{eqnarray}
    g_{i}(x) = \int^{1}_{0} dx \frac{\mathcal{N}_{i} (x)}{x(1-x) -r}\, log\frac{x(1-x)}{r}
\end{eqnarray}
where $\mathcal{N}_{h_{1,2,3}} = 2 x (1-x) -1$ and $\mathcal{N}_{A,a}(x) = 1$. As discussed before,
in the two-loop contribution as well 
we have not taken into account the SM Higgs contribution.

Moreover, in the present work, we also have one loop contribution mediated by the $U(1)_{B_{i}-L_{i}}$
gauge boson and has the following contribution to the $(g-2)_{\mu}$ 
\cite{Gninenko:2001hx, Baek:2001kca},
\begin{eqnarray}
    \Delta a^{Z^{\prime}}_{\mu} = \frac{g^{\prime\,2}}{8 \pi^{2}} \int^{1}_{0} d x \frac{2 x (1-x)^{2}}
    {(1-x)^{2} + r x}\,.
    \label{gb-to-muon-g-2}
\end{eqnarray}
As evidenced in the next section, the interplay of the different kind of contributions allows us to reproduce the $(g-2)$ anomaly in a relatively large parameter space.

\section{Results\,: Allowed parameter regions in different planes}
\label{result}

We are now going to display allowed regions among the parameters by performing a scan over the ranges summarized below. Each parameter assignation has been subject to bounds from DM relic density, $(g-2)$ as well as the theoretical bounds mentioned in in Section \ref{unitarity-bfb}. Only the model points passing all these constraints have been retained. For what concerns the 2HDM configurations, we have just focussed on the Type-II and Type-X scenarios.
Notice that in the Type-II scenario, we have imposed a priori the bound $M_{H^{\pm}} >800$ GeV to comply automatically with the bounds on $b \rightarrow s \gamma$ \cite{1503.01789}.
The Ref. \cite{1503.01789} also mentioned that for $\tan\beta < 2$, the bound will change, so in our study 
we have considered $\tan\beta > 2.5$ throughout the paper. We will discuss in detail the scanning plots
that will exhibit the correlations among the parameters and will deliver us a broader idea about the choice of the model
parameters. Additionally, in the case of Type-X 2HDM, we are about to see that the parameters are more relaxed 
but it does not provide us with the effective coupling between the axion and the gluons 
(see section \ref{axion-part}). Therefore, we choose to study the Type-II (Type-X) 
2HDM scenario which serves our 
purpose of having axion DM. We have varied our model parameters in the following range,
\begin{eqnarray}
    && 2.5 \leq \tan\beta \leq 250\,,\,\, 65 \leq M_{h_{1}}\,\,{\rm [GeV]} \leq 1500\,,\,\,
    125 \leq M_{h_{3}} \,\,{\rm [GeV]} \leq 1500\,, \nonumber \\
   && 10^{11} \leq v_{\phi_{1}}\,\,{\rm [GeV]} \leq 10^{16}\,,\,\,10^{-6} \leq g^{\prime} \leq 10^{-2}\,,
    \,\, 10^{-3} \leq M_{Z^{\prime}}\,\,{\rm [GeV]} \leq 1\,,\,\, \nonumber \\
    && 800\,(80) \leq M_{H^{\pm}}\,\,{\rm [GeV]} \leq 1500\,,\,\, 10^{5} \leq M_{\phi_{2}}\,\,{\rm GeV} \leq 10^{6}\,,\,\,
 \beta - \alpha_{1} \simeq \frac{\pi}{2}\,,
  \nonumber \\
    &&
  10^{-4} \leq \alpha_{2,3} \leq 10^{-1}\,, \,\, 
    10^{-4} \leq \lambda_{1} \leq 4 \pi\,.
\end{eqnarray}
We have considered $h_{2}$ as the SM Higgs and kept its mass fixed at $M_{h_{2}} = 125.5$ GeV.
The mass for the CP-odd scalars $A, a$ depends on the vevs, quartic couplings ($\lambda_{1,2}$), 
scalars mass and mixing angles and have been evaluated automatically during the scanning of the parameters.
The parameter $\mu$ has been measured using the quartic coupling $\lambda_{1}$ and the other Higgses masses and mixing angles.
 
\begin{figure}[h!]
\centering
\includegraphics[angle=0,height=7.5cm,width=7.5cm]{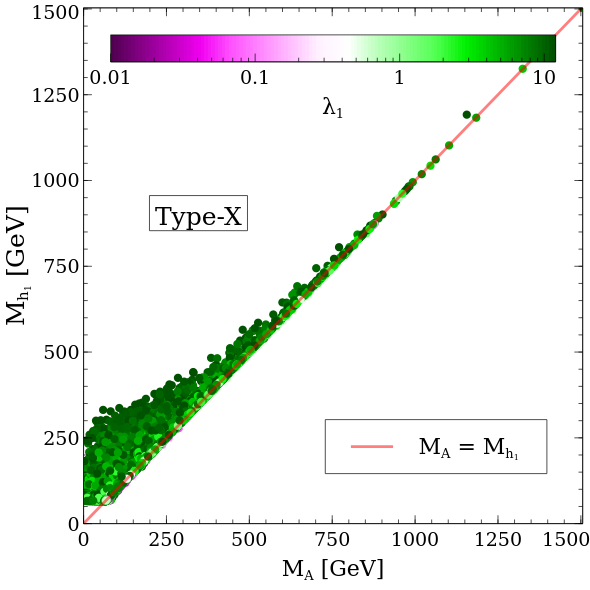}
\includegraphics[angle=0,height=7.5cm,width=7.5cm]{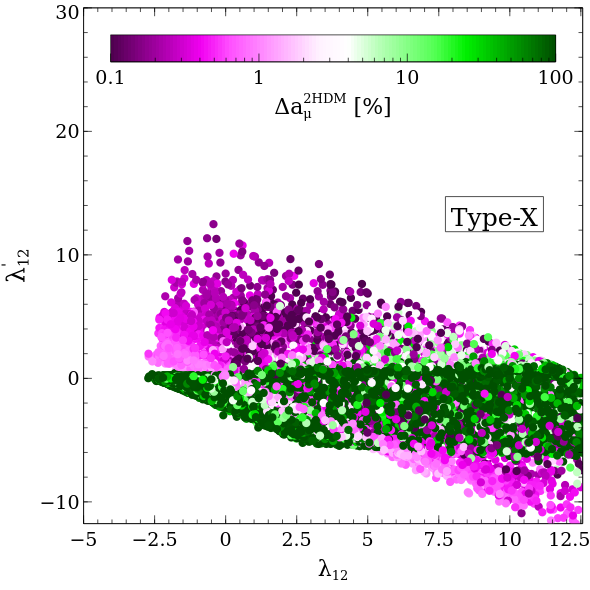} \\
\includegraphics[angle=0,height=7.5cm,width=7.5cm]{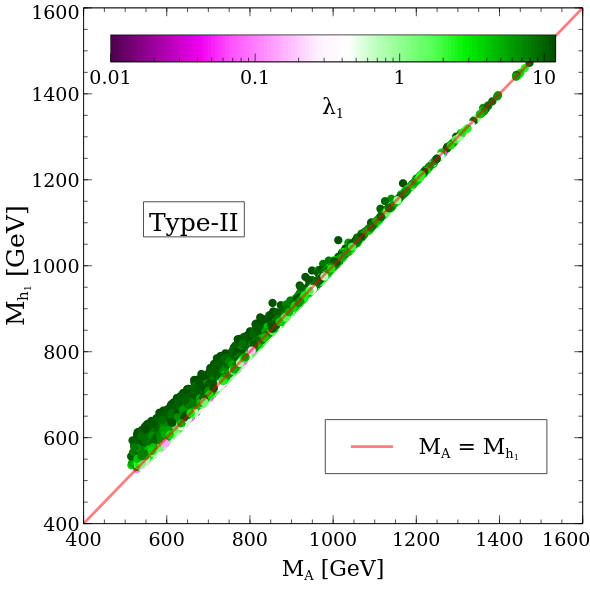}
\includegraphics[angle=0,height=7.5cm,width=7.5cm]{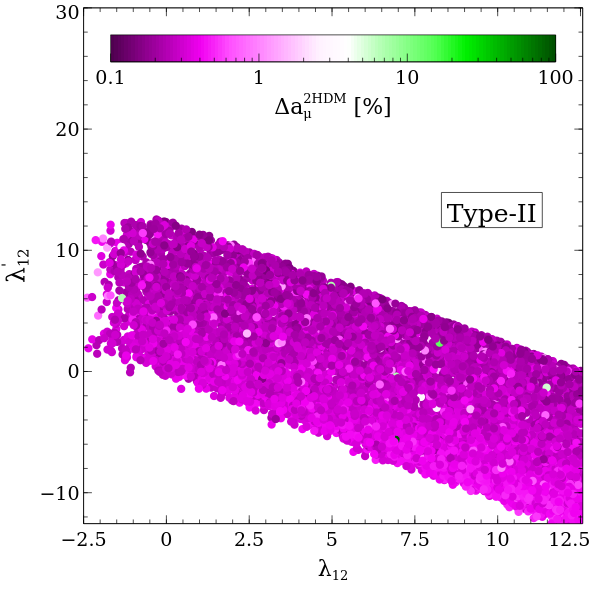} 
\caption{LP: Scatter plot in the $M_{A}-M_{h_{1}}$ plane after satisfying all the relevant bounds where
the colour bar corresponds to the different values of quartic coupling $\lambda_{1}$ associated with 
the Higgs doublet $H_{1}$. RP: Scatter plot in the $\lambda_{12}-\lambda^{\prime}_{12}$ plane and the colour
bar represents the contribution of the Higgs sector to muon ($g-2$) in percentage.} 
\label{scat-1}
\end{figure}

In Fig. \ref{scat-1}, we have shown scatter plots in the $M_{A} - M_{h_1}$ and 
$\lambda_{12} - \lambda^{\prime}_{12}$ planes after satisfying DM total relic density, muon $(g-2)$
and the perturbativity and bound from the below conditions on the quartic couplings. In the upper panel,
we have shown the result for the Type-X 2HDM model and in the lower panel we have shown it for 
Type-II 2HDM.
In scanning the parameters,
we have used the usual misalignment mechanism in 2HDM and $h_{2}$ as the SM like Higgs with mass 
125.5 Gev. Let us first discuss the upper panel and the lower panel will be followed thereafter.
We see a sharp correlation between $M_{A}$ and $M_{h_1}$ because the scatter plot follow the 
$M_{A} = M_{h_1}$ line represented by the red line. In general, the quartic coupling $\lambda_{1}$ depends on
the masses of the scalars and the $\tan\beta$ for $\alpha_{2,3} \sim 0$, so in the misalignment limit 
it can be expressed as,
\begin{eqnarray}
    \lambda_{1} \simeq \frac{1}{2 v^2_{1}} \left[ \left( M^2_{h_{1}} - M^2_{A} \right) \sin^{2}\beta  + 
    \cos^{2}\beta M^2_{h_{2}} \right]\,.
\end{eqnarray}
For larger values of $\tan\beta$, we have $\cos\beta \ll 1$ and $\sin\beta \sim 1$, then the $\lambda_{1}$
will be smaller than $4\pi$ only when we have $M_{A} \sim M_{h_1}$ {\it i.e} represented by the red 
diagonal line. In the other regime when we have $\sin\beta \ll 1$ and $\cos\beta \sim 1$, then we can go for
$M_{h_{1}} > M_{A}$ because it is suppressed by $\sin\beta$ and at the same time $v_{1} \sim v$, so the coupling 
additional suppression will happen due to the larger values of $v_{1}$ which is not possible for $\tan\beta \gg 1$.
The points that are outside the $M_{A} \sim M_{h_1}$ line correspond to the larger values of $\lambda_{1}$.
On the other hand, the points that are on the red line can take $\lambda_{1} < 0.1$ because of the 
possibility of the exact cancellation among the masses. In the RP, we have shown the scatter plot in the 
$\lambda_{12}-\lambda^{\prime}_{12}$ plane. We see that $\lambda_{12}$ and $\lambda^{\prime}_{12}$
follow an anticorrelation. This happens because we have used the condition 
on $\lambda_{12}$ and $\lambda^{\prime}_{12}$ {\it i.e.} $\lambda_{12}+ \lambda^{\prime}_{12}  < 4 \pi$ which makes the
plot looks like this. In particular if we look at $\lambda_{12} = 4 \pi$ then 
$\lambda^{\prime}_{12} = 0$ and vice versa and this follows all over the points. In the plots, we have
also shown the colour variation for each point which represents the scalar sector contribution
to muon $(g-2)$. The more green points represents $\lambda^{\prime}_{12} < 0$ and $\tan\beta$ larger
values which corresponds to $M_{A} < M_{H^{\pm}}$ and hence contribute in the muon ($g-2$) more.
Additionally, each point also satisfies the condition $\lambda_{12} + \lambda^{\prime}_{12} > - 2 \sqrt{\lambda_{1} \lambda_{2}}$ which keeps the possibility that their sum can be negative as well
and not exactly zero. In the lower panel, we have shown the same plots but for the Type-II
2HDM model. Here one of the main constraints comes from the $b \rightarrow s \gamma$ which 
demand charged Higgs mass $M_{H^{\pm}} > 800$ GeV. Due to such high values, of $M_{H^{\pm}}$, we also need $M_{A}$ around that order
otherwise $\lambda_{12}, \lambda^{\prime}_{12}$ would violate pertubativity.
Due to such a high value of $M_{H^{\pm}}$, we see a very sharp correlation and the starting value for 
$M_{A}$ is around 500 GeV. In the RP of the lower panel, we have shown the scatter plot in $\lambda_{12}-
\lambda^{\prime}_{12}$ plane and the colour bar show the contribution from the scalar sector in $(g-2)_{\mu}$.
We can see from the colour bar that most of the contribution comes from the gauge boson-mediated process.

\begin{figure}[h!]
\centering
\includegraphics[angle=0,height=7.5cm,width=7.5cm]{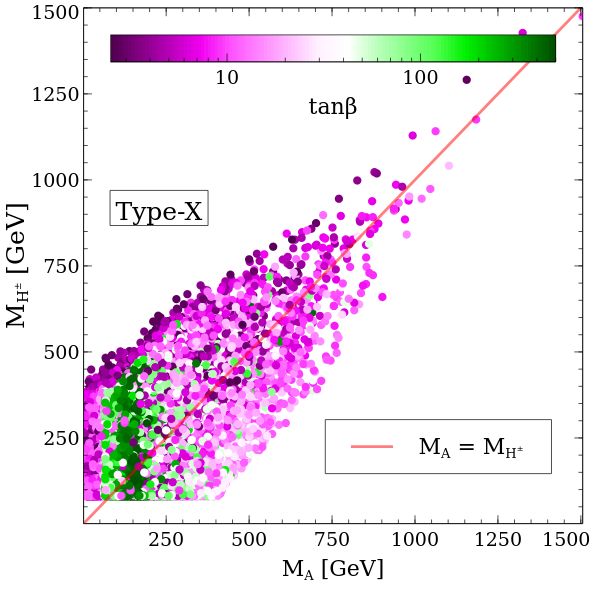}
\includegraphics[angle=0,height=7.5cm,width=7.5cm]{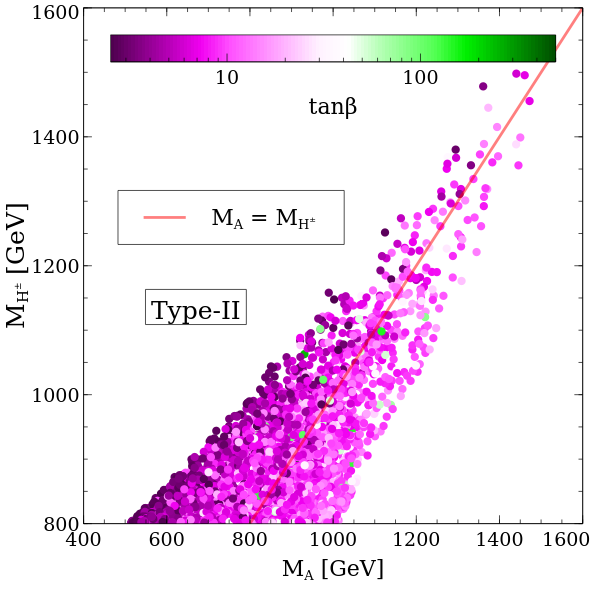}
\caption{LP shows the scatter plot in the $M_{A}-M_{H^{\pm}}$ plane for Type-X 2HDM where the colour bar shows the variation w.r.t $\tan\beta$. In the RP we have also shown the variation in the same plane with
the same colour variation but for Type-II 2HDM.} 
\label{scat-new}
\end{figure}

In the LP and RP of Fig. \ref{scat-new}, we have shown scatter plots in the $M_{A}-M_{H^{\pm}}$ for 
for Type-X and Type-II 2HDM models and in both plots the colour variation shows different values 
of $\tan\beta$. The allowed region is mostly obtained by the perturbativity and the bound from
the below conditions on the quartic couplings. In the misalignment limit and $\alpha_{2,3} \rightarrow 0$,
the quartic couplings $\lambda_{12}, \lambda^{\prime}_{12}$ can be expressed in terms of the masses
as follows,
\begin{eqnarray}
    \lambda_{12} \simeq \frac{2 M^2_{H^{\pm}} - \left( M^2_{A} + M^2_{h_{1}} - M^2_{h_2} \right) }{v^{2}}\,,\,\,
    \lambda^{\prime}_{12} \simeq \frac{2}{v^{2}} \left( M^2_{A} - M^2_{H^{\pm}} \right) \,.
\end{eqnarray}
From the above expressions, we can see that the masses of the scalars can not be very much apart
otherwise, it will conflict with the perturbativity bound. In Fig. \ref{scat-1} we have shown
a sharp correlation between $M_{A}$ and $M_{h_1}$ which will ensure also a sharp correlation 
with the charged Higgs mass as seen in both the LP and RP. In the LP for Type-X 2HDM, we don't have
strong constraints on the charged Higgs mass, hence small mass ranges of $M_{A}$ and $M_{H^{\pm}}$ are allowed.
The colour variation on the points implies the different values of $\tan\beta$. The muon ($g-2$) depends on 
$\tan\beta$ and $M_{A}$ generally as $\frac{\tan^{2}\beta }{M^{2}_{A}}$ as can be seen in Section \ref{muon-g-2}. 
So for the higher values of $M_{A}$ we also need higher values of $\tan\beta$ but we have taken maximum 
$\tan\beta = 500$. In general, higher values of $\tan\beta$ will impact the 
$\lambda_{1}$ perturbativity bound and also the Yukawa coupling perturbativity bound 
for the b-quark ($\tau$-lepton) for Type-II (Type-X) 2HDM model
puts upper bound on $\tan\beta < 206\,\, (\tan\beta < 500)$.
In the RP, we have shown it for the Type-II
2HDM case and we have strong constraints on the charged Higgs mass from $b \rightarrow s \gamma$
{\it i.e.} $M_{H^{\pm}} > 800$ GeV. Therefore, we see allowed range also starts at higher values of 
$M_{A}$. Therefore, for this case, we do not have a dominant contribution
from the scalar sector in ($g-2$) if a very accidental cancellation does not happen 
among the masses which makes $\lambda_{1}$ perturbative for higher values of
$\tan\beta$ as well. The dependence of Type-II and Type-X 2HDM models on muon ($g-2$) will be more clear in the
later part.

\begin{figure}[h!]
\centering
\includegraphics[angle=0,height=7.5cm,width=7.5cm]{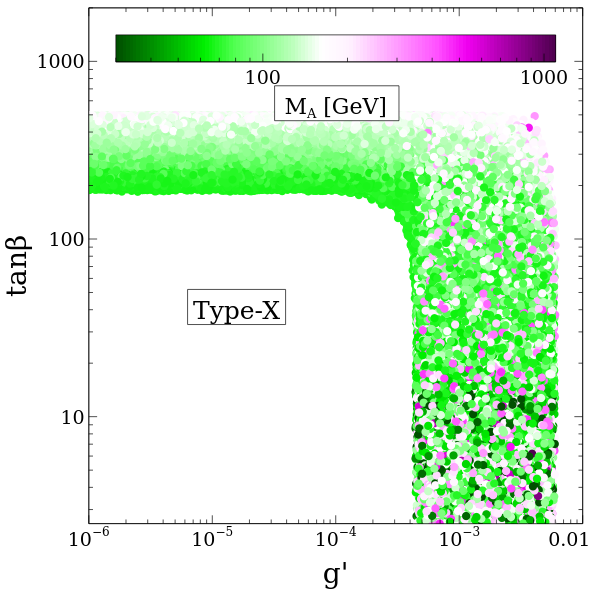}
\includegraphics[angle=0,height=7.5cm,width=7.5cm]{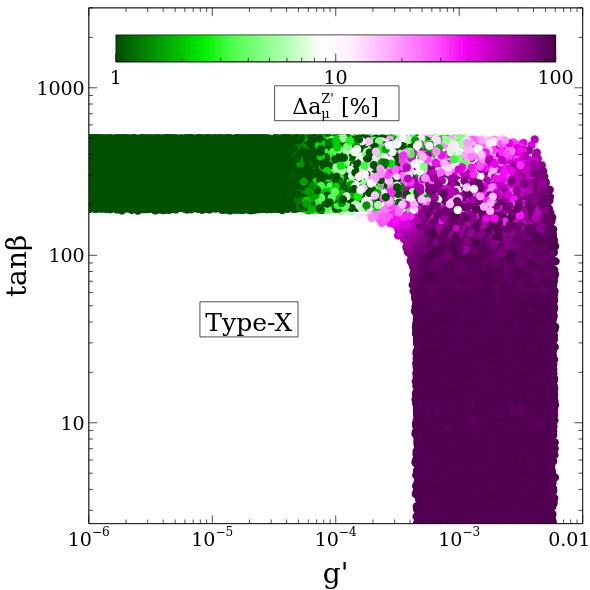}\\
\includegraphics[angle=0,height=7.5cm,width=7.5cm]{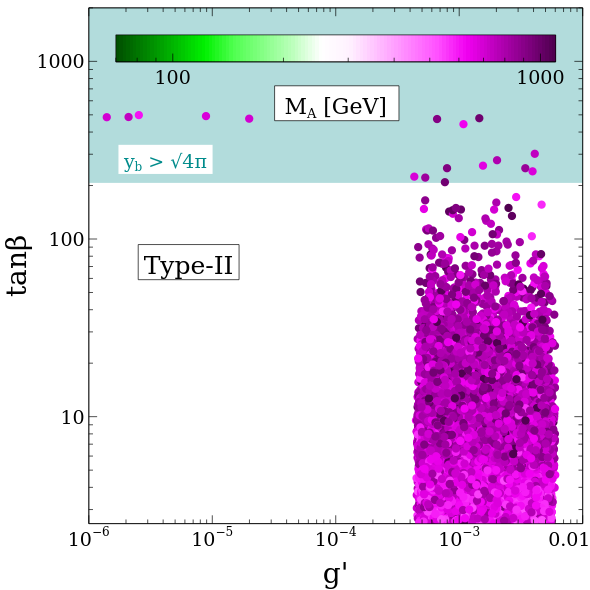}
\includegraphics[angle=0,height=7.5cm,width=7.5cm]{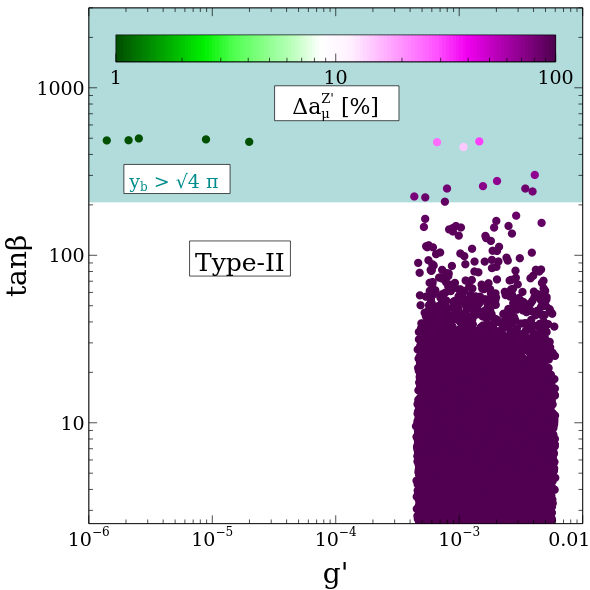}
\caption{LP shows the scatter plot in the $g^{\prime}-\tan\beta$ plane where the colour bar shows the variation w.r.t the mass of the CP odd Higgs $A$. The RP also shows the variation in the same plane but 
 the colour bar represents the $Z^{\prime}$ contribution to $(g-2)_{\mu}$.} 
\label{scat-2}
\end{figure}

In the LP and RP of Fig. \ref{scat-2}, we have shown scatter plots in the $g^{\prime} - \tan\beta$ plane
where the colour bars represent the variation for the CP-odd scalar mass and the gauge boson 
contribution to muon $(g-2)$ in percentage. Let us first discuss the upper panel which is for Type-X
2HDM model.
The general contribution from the Higgs sector
to $(g-2)_{\mu}$ can be parametrised as $\Delta a^{2HDM}_{\mu} \propto \frac{\tan^{2}\beta}{M^2_{A}}$. As we have
seen earlier in Fig. \ref{scat-1} that $M_{A} < 100$ GeV is allowed for smaller values of $\tan\beta$
due to the perturbativity limit on $\lambda_1$. As explained before $(g-2)_{\mu}$ is inversely proportional to the 
mass square $M^2_{A}$ and linearly to the square of $\tan\beta$. Therefore, $g^{\prime} < 4 \times 10^{-4}$
and $\tan\beta < 200$, we do not have any points because $(g-2)_{\mu}$ can not be produced 
in sufficient. Once, we have a sufficient contribution to $g-2$
for the parameter space $60 < M_{A} \,\,{\rm GeV} < 300$ and $\tan\beta > 200$, then the most of the contribution to $(g-2)_{\mu}$ anomaly
comes from the Higgs sector. On the other hand when $g^{\prime} > 4 \times 10^{-4}$ then we can produce the 
total $g-2$ contribution solely from the $Z^{\prime}$, therefore for those parameter spaces lower values of $\tan\beta$ 
as well as the higher values of $M_{A}$ are also allowed. On the other hand in the right panel, we have shown the
colour variation w.r.t the $U(1)_{X}$ gauge boson contribution to $(g-2)_{\mu}$. 
As can be seen from Eq. \ref{gb-to-muon-g-2}, gauge boson contribution to muon $g-2$ is proportional to $g^{\prime\,2}$. This kind of behaviour can be easily seen from the colour variation like for $g^{\prime} < 4\times 10^{-4}$, the gauge boson contribution is subdominant but for the opposite regime, it is the dominant one. The higher values of $g^{\prime}$, represented by the region $g^{\prime} > 6 \times 10^{-3}$, are also ruled out because that region will produce a more positive contribution to muon $(g-2)$ anomaly. The green points represent the 
dominant contribution from the scalar sector whereas the magenta points represent the dominant
contribution from the gauge boson sector $Z^{\prime}$. In the lower panel, we have shown the
same kind of plots but for the Type-II 2HDM model. Since here we have to obey 
$M_{H^{\pm}} > 800$ GeV then we can see larger values of $M_{A}$ are allowed due to the pertubativity
bound. In the LP, we can see a few points at lower $g^{\prime}$ those points correspond to 
$M_{A} = M_{h_1}$ and small value of $M_{h_3}$, then those points can pass the perturbativity bound 
for a high value of $\tan\beta$ value as well. Therefore, those points can give us a full contribution
to muon ($g-2$) from the scalar sector only. Although those points fall in the cyan region which corresponds to the 
perturbative violation for the of the $b-$quark Yukawa coupling.
In the RP, we can see the distinction between the points which distinguish the contribution from the scalar and gauge boson sectors.
Again the points that represent dominant contribution from the scalar sector are ruled out by the perturbativity bound on the 
$b-$quark Yukawa coupling.

\begin{figure}[h!]
\centering
\includegraphics[angle=0,height=7.5cm,width=7.5cm]{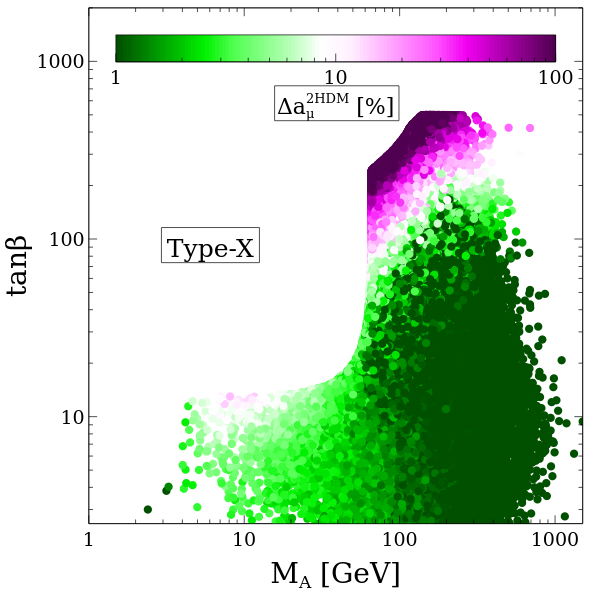}
\includegraphics[angle=0,height=7.5cm,width=7.5cm]{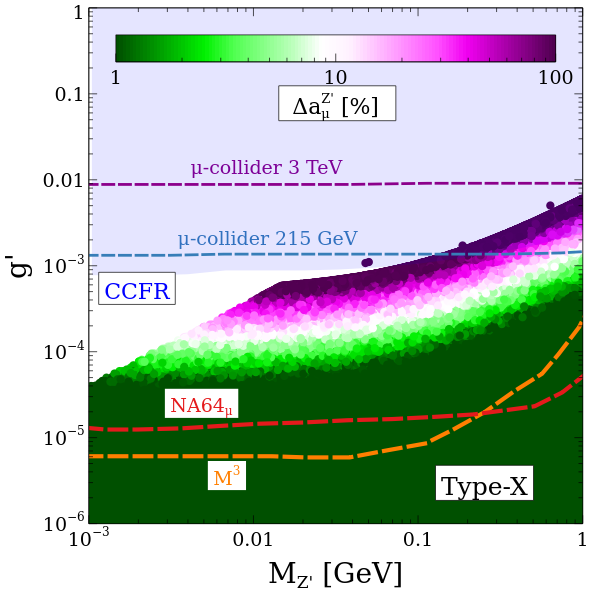}\\
\includegraphics[angle=0,height=7.5cm,width=7.5cm]{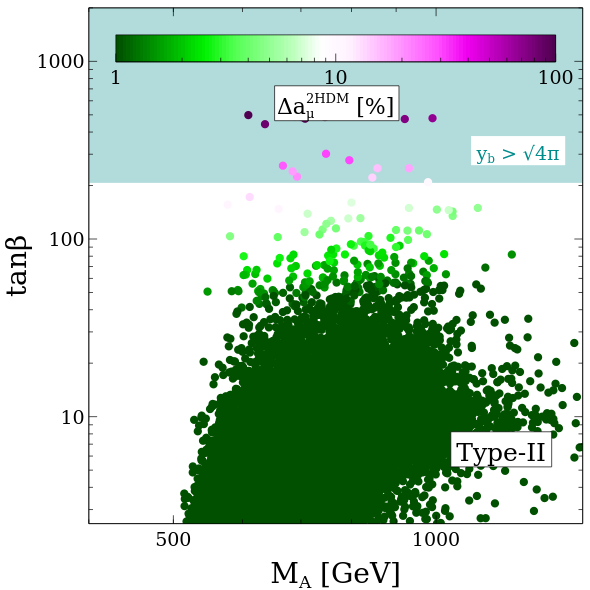}
\includegraphics[angle=0,height=7.5cm,width=7.5cm]{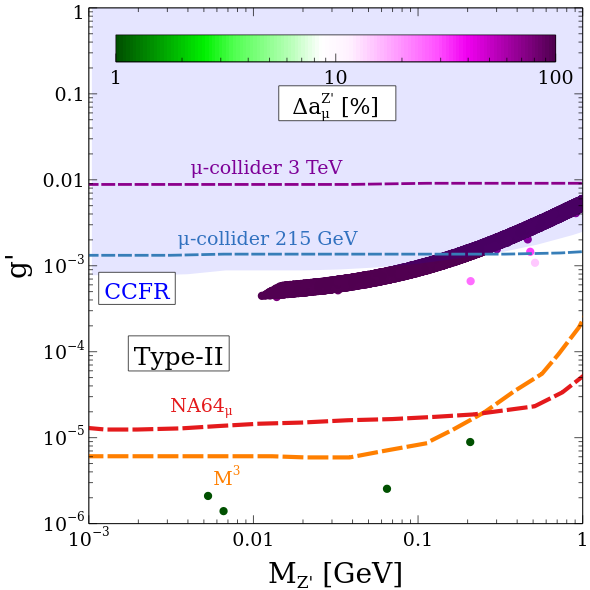}
\caption{LP and RP show the scatter plots in the $M_{A}-\tan\beta$ and $M_{Z^{\prime}}-g^{\prime}$ planes,
respectively. The colour variation in both plots represents the scalar sector contribution and
gauge boson contribution to the muon ($g-2$) in percentage.} 
\label{scat-3}
\end{figure}

In the the LP and RP of Fig. \ref{scat-3}, we have shown the scatter plots in the $M_{A}-\tan\beta$
and $M_{Z^{\prime}}-g^{\prime}$ planes where the upper panel corresponds to the Type-X 2HDM and
the lower panel corresponds to the Type-II 2HDM model. 
We first discuss the upper panel then we discuss the lower panel.
As discussed LP of Fig. \ref{scat-3} shows the allowed region in the $M_{A} - \tan\beta$ plane where the 
colour variation represents the Higgs sector contribution to muon $(g-2)$. As can be seen,
for $\tan\beta < 60$ the Higgs sector contribution to $(g-2)_{\mu}$ is less than 10 \%. Once we go to larger values of $\tan\beta$, say $\tan\beta > 100$, we see the magenta points which represent more than $50\,\%$ contribution to $(g-2)_{\mu}$. The colour variation in the 
figure also clearly explains that for a particular value of $M_{A}$, if we go for higher values 
of $\tan\beta$ then we gradually move towards the magenta points {\it i.e.} more contribution 
to $(g-2)_{\mu}$ from the Higgs sector. 
The figure also explains that if one wants that the muon $(g-2)$ anomaly is accounted for the Higgs sector, the viable parameter space is very narrow as it is represented only by the deep magenta points. The presence of the extra gauge boson is essential to enlarge the parameter space. We notice additionally, that the distribution of the points complying with $(g-2)$ has extended for $M_A < 62.5$ 
as well. 
These points can conflict with the LHC bound coming from $h_{2} \rightarrow AA$ decay but needs 
proper treatment which limits our further discussion. 
On the other hand in the RP, we have shown the scatter plot in the $M_{Z^{\prime}} - g^{\prime}$
plane. Here, the deep magenta region can only explain the $(g-2)_{\mu}$ anomaly
fully if we consider only the gauge boson contribution. But because of the presence of the Higgs sector the region expands for lower values $g^{\prime}$ as well represented by the 
green points. The region below $M_{Z^{\prime}} < 0.01$ GeV and 
$ 4 \times 10^{-5} < {g^{\prime}}< 8 \times 10^{-4}$ have no points because those points
make $\lambda_{\phi_2} > 4 \pi$. A part of the allowed region in the 
$M_{Z^{\prime}} - g^{\prime}$ plane has already explored by the CCFR neutrino 
trident experiment \cite{Altmannshofer:2014pba}, represented by the blue shaded region in the plot. Moreover,
the remaining parameter space will be explored through different experiments like 
$NA64_{\mu}$ at CERN \cite{Gninenko:2014pea}, 
$M^{3}$ at Fermi lab \cite{Kahn:2018cqs} and even at the proposed muon collider \cite{MuonCollider:2022xlm}. 
All the future bounds are represented 
by the dashed lines and labelled according to the associated experiment name.
In the lower panel, the figures are in the same plane and also with the same kind of 
color variation for the Type-II 2HDM case.
As we have seen before, scalar sector contribution to muon (g-2) is subdominant so the full contribution 
comes from the gauge boson sector which can be seen from both left and right plots. There are a few points
which can give $100\%$ are the points which satisfy $M_{A} = M_{h_1}$ and $M_{h_3}$ value is small say near
to the SM Higgs mass. But those few points are already contradictory due to the perturbative violation of 
the $b-$quark Yukawa coupling.

\section{Explnation of W-boson mass observed at CDF-II}
\label{w-mass}
Recently, the CDF-II collaboration has reported the larger values of W-boson mass based on their $8.8\,\,fb^{-1}$
data of $\sqrt{s} = 1.96$ TeV run of $p\bar{p}$ collision. They have obtained the
W-boson mass \cite{CDF:2022hxs},
\begin{eqnarray}
    M^{CDF-II}_{W} = 80.4335 \pm 0.0094\,\, {\rm GeV}
\end{eqnarray}
while there is also a prediction for the W-boson mass from other collaborations 
like LEP \cite{CDF:2013dpa}, ATLAS \cite{ATLAS:2017rzl},
LHCb \cite{LHCb:2021bjt}, Tevatron \cite{CDF:2013bqv, CDF:2022hxs} and can be summarised as,
\begin{eqnarray}
    M_{W} &=& 80.4112 \pm 0.0076\,\, {\rm GeV},\,\,\,\,{\rm LHC+LEP+Tevatron}\,,\nonumber \\
    M_{W} &=& 80.3790 \pm 0.0120\,\, {\rm GeV},\,\,\,\,{\rm PDG 2020}\,.
\end{eqnarray}
The values disagree with each other with quite a high statistical significance as the CDF-II disagrees
with the SM values at the significance of $7\sigma$. 
We need beyond standard model contribution to address the increased value  
of the W-boson mass at the CDF-II detector. The recent reanalysis of ATLAS data 
collected in 2011
for the 7 TeV run corresponding to $4.6 fb^{-1}$ integrated luminosity 
using the state-of-the-art analysis methods 
estimated the W-boson mass which is in agreement with the SM predicted value \cite{ATLAS:2023fsi}.
Therefore, it is very early
to conclude that the discrepancy observed at the CDF-II measurement 
is due to the BSM physics, still, investigate  
the possibility of the present model to explain the discrepancy at CDF-II as well as the 
SM predicted value. 
In the present work, we have additional particles which can give extra contributions
to the W-boson mass and hence potentially account for the experimental anomaly. In order to investigate quantitatively such possibility, 
we have calculated the oblique parameters $S, T$ and $U$ by following the 
Ref. \cite{Grimus:2008nb} for the present work which also coincides with 
the result obtained in Refs. \cite{Muhlleitner:2016mzt, Arhrib:2018qmw} in particular limit. 
The deviation of the mass of the W-boson, with respect to the SM prediction, can be expressed in terms of the oblique parameters as \cite{Grimus:2008nb}:
\begin{eqnarray}
    M^2_{W} = M^{2}_{W}\bigg|_{SM} \times \left[ 1 + \frac{\alpha}{c^2_{w} - s^2_{w}} \left( -\frac{1}{2} S + c^2_{w} T
    + \frac{c^2_{w} - s^2_{w} }{4 s^2_{w}} U \right) \right]\,,
\end{eqnarray}
where $s^2_{w} = \sin^{2} \theta_{w} \simeq 0.23$ and $\alpha$ is the fine structure constant at the scale of 
Z-boson pole mass. From the above expression, it is clear that if the oblique parameters $S, T$ and $U$ are small
or zero then the W-boson mass coincides with the SM value.

\begin{figure}[h!]
\centering
\includegraphics[angle=0,height=6.0cm,width=5.5cm]{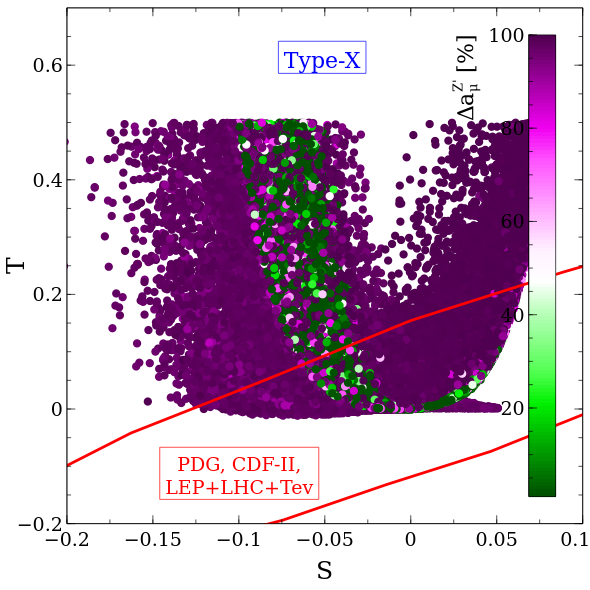}
\includegraphics[angle=0,height=6.0cm,width=5.5cm]{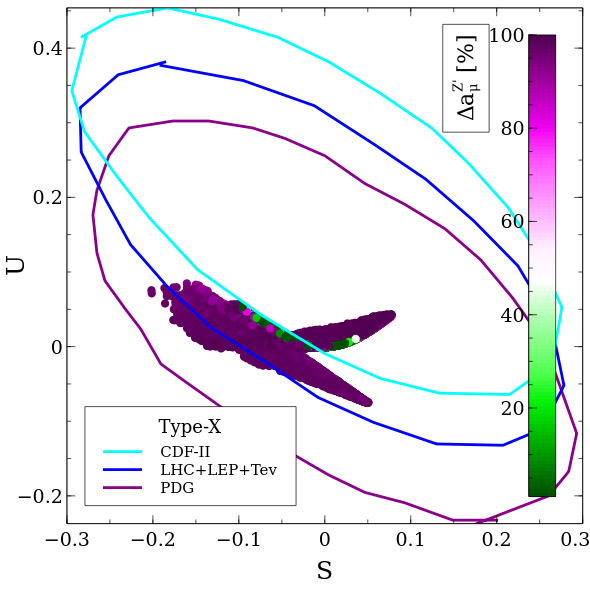}
\includegraphics[angle=0,height=6.0cm,width=5.5cm]{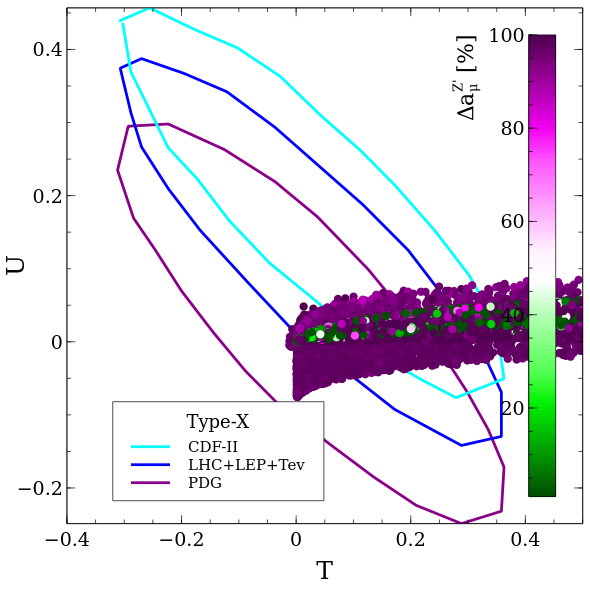} \\
\includegraphics[angle=0,height=6.0cm,width=5.5cm]{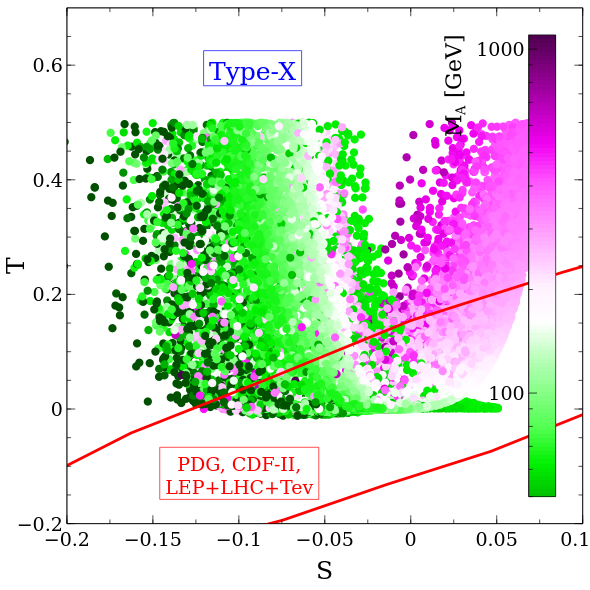}
\includegraphics[angle=0,height=6.0cm,width=5.5cm]{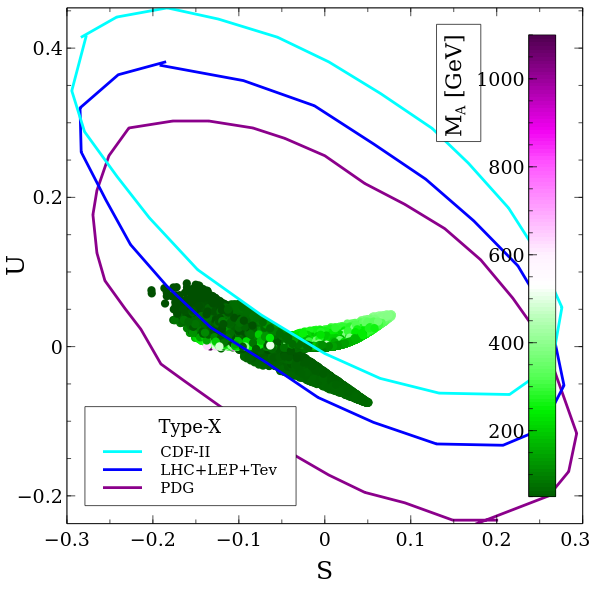}
\includegraphics[angle=0,height=6.0cm,width=5.5cm]{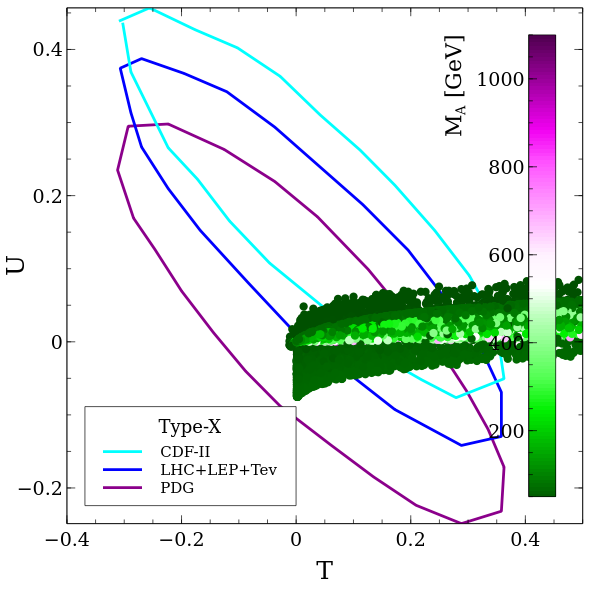}
\caption{Scatter plots in $S-T$, $S-U$ and $T-U$ planes after satisfying total DM relic density, muon $(g-2)$,
pertubativity and Unitarity bounds. In the upper panel, the colour bars show the variation of the $Z^{\prime}$
contribution to muon $(g-2)$ in percentage whereas the lower panel shows the variation w.r.t mass of the CP
odd Higgs $A$. All the figures are generated by using a Type-X 2HDM kind of model.} 
\label{scat-4}
\end{figure}

\begin{figure}[h!]
\centering
\includegraphics[angle=0,height=6.0cm,width=5.5cm]{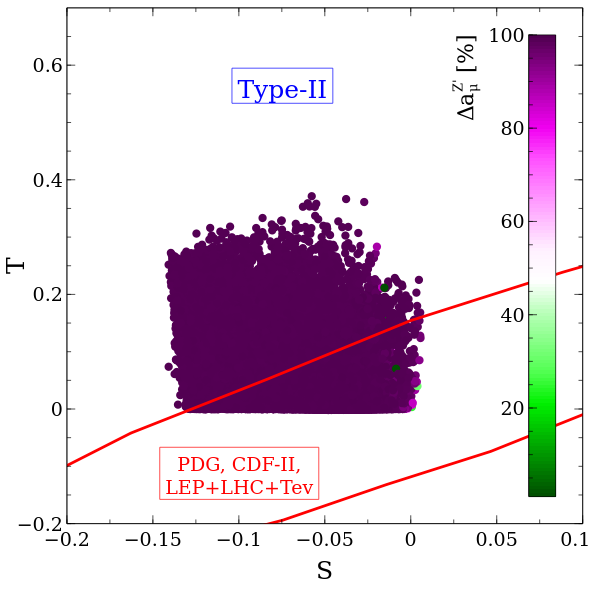}
\includegraphics[angle=0,height=6.0cm,width=5.5cm]{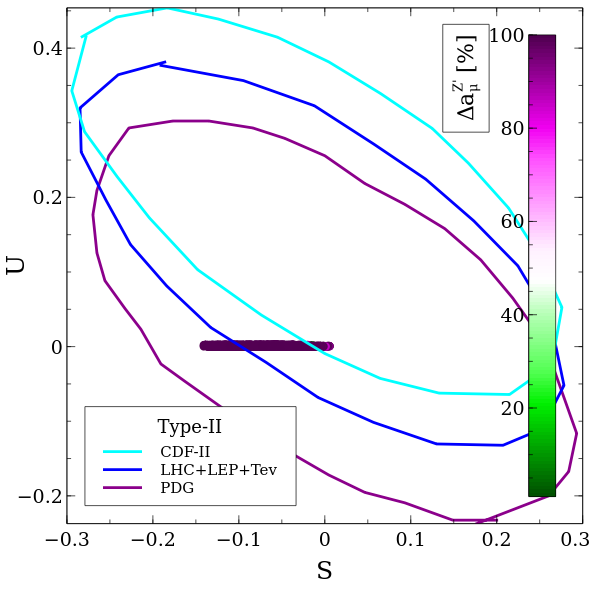}
\includegraphics[angle=0,height=6.0cm,width=5.5cm]{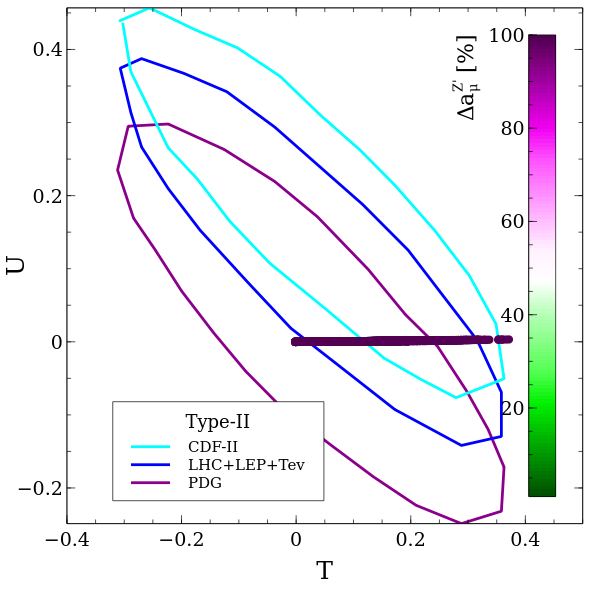} \\
\includegraphics[angle=0,height=6.0cm,width=5.5cm]{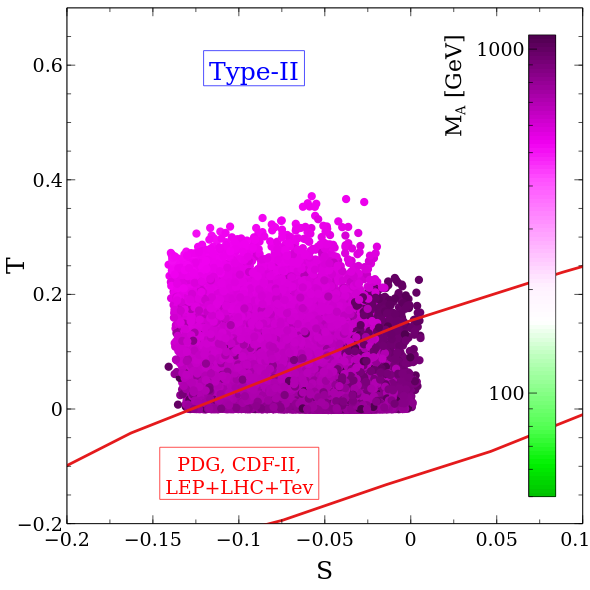}
\includegraphics[angle=0,height=6.0cm,width=5.5cm]{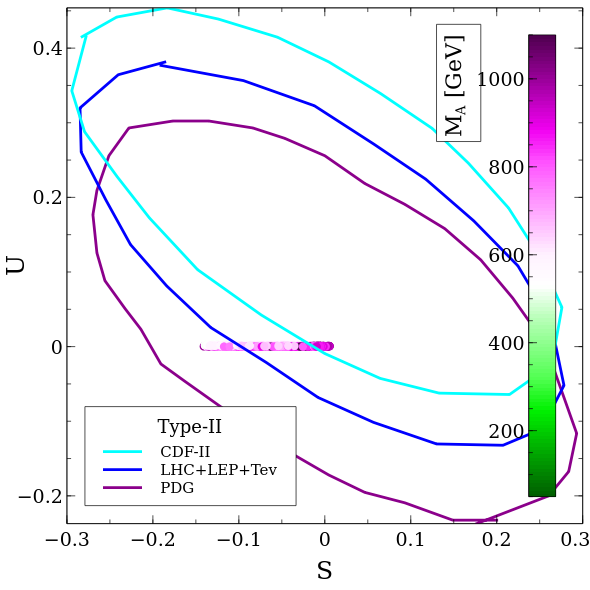}
\includegraphics[angle=0,height=6.0cm,width=5.5cm]{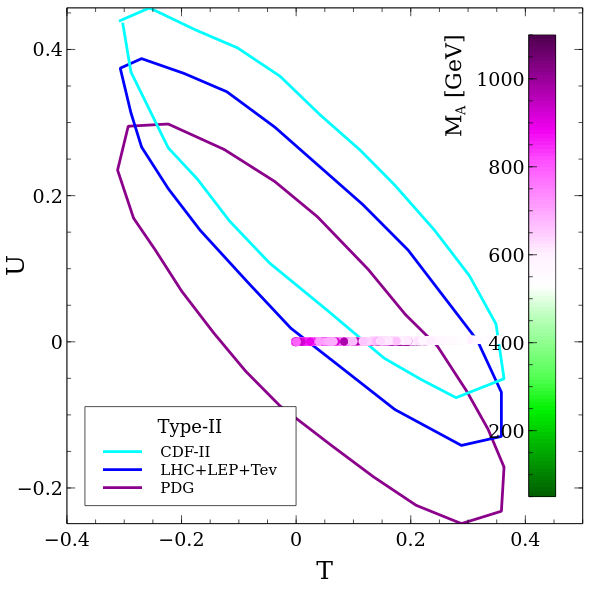}
\caption{Scatter plots in $S-T$, $S-U$ and $T-U$ planes after satisfying total DM relic density, muon $(g-2)$,
pertubativity and Unitarity bounds. In the upper panel, the colour bars show the variation of the $Z^{\prime}$
contribution to muon $(g-2)$ in percentage whereas the lower panel shows the variation w.r.t mass of the CP
odd Higgs $A$. All the plots have been generated by using a Type-II kind of 2HDM model.} 
\label{scat-5}
\end{figure}

In Fig. \ref{scat-4}, we have shown scatter plots among the oblique parameters plane
namely $S-T$, $S-U$ and $T-U$ in the context of the Type-X 2HDM model. In the upper panel, we have shown the percentage of
gauge boson contribution to $(g-2)_{\mu}$ whereas in the lower panel, we have shown the 
colour variation in terms of the mass of the CP-odd scalar $A$. Through these plots, we have 
tried to show that our model can explain all the measured values of the W-boson mass from ATLAS
to CDF-II. The contour plots in different oblique parameters plane which fit
the W-boson mass from different measurements namely PDG, CDF-II and
LEP+LHC+Tev have been borrowed from Ref. \cite{Asadi:2022xiy}. 
In the first plot of the upper panel
we have shown the allowed region in the $S-T$ plane and the red contour plot represents the 
allowed region to explain the W-boson mass from different measurements. We can see a substantial amount of points 
within the red contour region.  
Similarly, we have shown the allowed region in the $S-U$ and $T-U$ planes 
where the allowed regions from different experimental measurements of W-boson mass
are represented as cyan, blue and magenta contour plots. We can see in both the plots we have common overlap regions which can explain the W-boson mass simultaneously from all the measurements. 
In generating the plots, we have considered $T< 0.5$
because beyond these values, there will be more contribution to the $W-boson$ mass and have already been 
ruled out from experiments. In the lower panel, we have shown the colour variation in terms 
of the mass values $M_{A}$. We can see that both lower and higher values of $M_{A}$ can
explain the W-boson mass. Finally, in Fig. \ref{scat-5}, we have also shown the region which can explain the 
W-boson mass from the different measurements in the Type-II 2HDM model. Here we need higher values of $M_{A}$ and the other
mass scales, so we have less beyond SM effects on the $S, T, U$ parameters which can be easily seen from the 
figures. All the points represent subdominant contribution to $(g-2)_{\mu}$ from the scalar sector so $100\%$ contribution comes from the 
gauge sector. The variation of the oblique parameters slightly differs from the SM value which is
$S,T,U = 0$.

\section{Summary and Conclusion}
\label{conclusion}

In this work, we have proposed and studied a framework aiming to address some of the most relevant puzzles which call for the existence of Physics beyond the Standard Model: Dark Matter, neutrino masses, $(g-2)_{\mu}$ and anomalous measure of the W-mass by CDF-II.  
In this context, we have taken 2HDM as the basis model and extended its particle 
content by two singlet scalars and three right-handed neutrinos. The gauge group has also been extended by an additional abelian gauge symmetry. Additionally, we have also introduced 
an additional global symmetry which can be resembled with the PQ symmetry and upon its breaking
we have the axion field. Among the two singlet scalars, one of them is charged under the 
$U(1)_{B_{i}-L_{i}}$ gauge group and its CP-odd component becomes the longitudinal part 
of the additional gauge boson and imparts its mass. On the other hand, the remaining CP-odd scalar coming from $\phi_1$ becomes an axion field when the global PQ symmetry breaks spontaneously due to $\phi_1$ vev. The axion field can be produced in the early universe 
by the misalignment mechanism and becomes a cold DM. In the case of Type-II configurations for the Yukawa couplings of the scalar sector, the axion DM can be identified with the QCD axion possibly solving the strong CP problem. Even if it cannot account for the solution of the latter puzzle, we have considered, in this work, the Type-X scenario as well as it appears to be more favourable for interpreting the $(g-2)$ anomaly and for neutrino masses. 
Another important beyond SM problem namely neutrino mass can also be addressed by the Type-I seesaw mechanism when
the scalars take the spontaneous vev. In particular, few elements in the RH neutrino mass 
matrix are suppressed
by the Planck mass therefore it is hard to generate the RH neutrino mass above 1 GeV for 
the Type-II 2HDM case but for the Type-X case RH neutrino mass can be 
generated as large as 1 TeV. Although, the aforementioned statement is true when we consider all the
elements in the right-handed neutrino mass matrix are the same they can be deviated by 
a few orders of magnitude obeying the neutrino oscillation data. 
We are also able to explain the muon ($g-2$)
anomaly in the present set-up. In our work, we have two contributions in the $(g-2)_{\mu}$, one coming
from the scalar sector and the other contribution coming from the additional gauge boson. 
As found in the case of Type-II 2HDM, due to strong constraint on the charge Higgs mass from 
$b \rightarrow s \gamma$ measurement
the scalar sector contribution to $(g-2)_{\mu}$ is negligible but for Type-X 2HDM
variant scalars can produce 100\% deficiency in $(g-2)_{\mu}$. Therefore, for Type-X we can have 
the freedom to choose gauge coupling and gauge boson mass which can give us 
higher values of vev and hence the possibility of obtaining TeV scale right-handed neutrinos.  
The gauge sector effect in $\Delta a_{\mu}$ is the same irrespective of Type-X or Type-II
variants of the 2HDM model. Additionally, due to the presence of additional particles, we can 
have an effect on the oblique parameters $S, T, U$ which can contribute to the W-boson mass. 
We have found that the allowed parameters after 
taking into account all the observables can also explain the CDF-II measurement in 
some part of the parameter space and the remaining part can address the SM 
predicted value {i.e.} subdominant BSM contribution.
Therefore, our model can explain both the data for the W-boson mass 
{i.e.} CDF-II and SM values of the W-bosn mass.
Finally, we conclude that our model can explain many beyond standard model drawbacks simultaneously
namely dark matter, neutrino mass, strong CP problem,
muon ($g-2$) and the potential discrepancy in the measurement of the W-boson mass. Additionally, 
the present work has the potential to explain the gravitation waves and inflation due to the
rich scalar sector which are left to pursue in future.

\section{Acknowledgements}
SK extends sincere thanks for the warm hospitality provided at the Department of Mathematical and Computer Sciences, Physical Sciences, and Earth Sciences, University of Messina during the visit 
when this project was initiated. SK acknowledges Laura Covi for useful discussions.
This work used the Scientific Compute Cluster at GWDG, the joint data center of Max
Planck Society for the Advancement of Science (MPG) and University of Göttingen.


\appendix

\section{Gauge Anomaly}
\label{gauge-anomaly}

Gauge anomaly conditions associated with the gauge symmetry are very crucial to make them zero otherwise the
theory would become non-renormalizable. For a generator $T^{a}$ of any gauge symmetry, the gauge anomaly term can be written as
\begin{eqnarray}
    \partial^{\mu} J^{a}_{\mu} = \left( \sum_{left} A(R_{l}) - \sum_{right} A(R_{r})  \right)
    \frac{g^{2}}{128 \pi^{2}} Tr\left[ T^{a}\{T^{b},T^{c}\} \right] F \tilde{F}\,,
\end{eqnarray}
where $A(R)$ depends on the group representation, for fundamental representation it is equal to 
$A(R)|_{funda} = 1$ and for adjoint or singlet representation is $A(R)|_{Adj\,\,or \,\,sing} = 0$.
In SM, the left-handed fermions are in fundamental representation whereas right-handed fermions are in singlet
representation, therefore, in the present work we mainly need to check 
the quantity $Tr\left[ T^{a}\{T^{b},T^{c}\} \right]$ if it vanishes or not for all possible combinations of
the gauge groups. It is clear that if the trace contains only one non-abelian generator then it is
trivially vanishes because of the traceless property of the non-abelian generator. Moreover, due to the vectorial
nature of the fermions under $U(1)_{X}$ gauge symmetry the gauge combinations 
$[SU(3)]^{2} \times U(1)_{X}$, $[U(1)_{X}]^{2} 
\times U(1)_{Y}$, $[U(1)_{X}]^{3}$ and $[Gravity]^{2} \times U(1)_{X}$ vanieshes. The non-zero
contributions include,
\begin{eqnarray}
    [SU(2)]^{2} \times U(1)_{X} &:& \frac{Tr[\sigma^{a} \sigma^{b}]}{2} \sum^{3}_{i=1} \left( 
    3 b_{i} + b^{\prime}_{i}\right) \nonumber \\
    U(1)_{Y}^{2} \times U(1)_{X} &:& \sum^{3}_{i=1} \left( -\frac{3}{2} b_{i} - \frac{1}{2} b^{\prime}_{i} \right)\,.
\end{eqnarray}
Non-zero contributions can be made zero by choosing,
\begin{eqnarray}
    \sum^{3}_{i=1} \left( 3 b_{i} + b^{\prime}_{i} \right) = 0\,.
\end{eqnarray}
Therefore, there are many ways we can choose $b_{i}, b^{\prime}_{i}$ (i = 1, 2, 2) to satisfy the above equation.
In this work, we have chosen the simplest combination which is $b_{i} = 0$, $b^{\prime}_{1} = 0$ 
and $b^{\prime}_{2} = -b^{\prime}_{3} = 1$
which is $U(1)_{L_{\mu} - L_{\tau}}$ abelian extension of the SM and has been extensively
studied in the literature. The more general kind of $U(1)$ gauge symmetry with different 
kinds of flavour combinations has been shown in \cite{Kownacki:2016pmx}.

\section{Bound on the quartic couplings}
\label{unitarity-bfb}
\begin{itemize}
    \item {\bf Bound from below:}
    In this part, we put the bound on the quartic couplings which are needed to prevent the
    total potential from becoming negative for very high values of the fields. In this part, we have assumed that $\lambda_{H_{1}\phi_{1}}, \lambda_{H_{2}\phi_{1}}, \lambda_{\phi_{1} \phi_{2}} 
    \sim 0$. As described in \cite{Klimenko:1984qx}, we can have two scenarios depending on
    $\lambda^{\prime}_{12} > 0 or < 0$.

{\bf Case I: $\lambda^{\prime}_{12} > 0$}
    
    \begin{eqnarray}
&&    \Omega_{1} 
=\left( \lambda_{\phi_{2}}, \lambda_{1}, \lambda_{2} \geq 0; \lambda_{H_{i}\phi_{2}} > 
-2 \sqrt{\lambda_{\phi_{2}} \lambda_{i} }; \lambda_{12} > - 2 \sqrt{\lambda_{1} \lambda_{2}}; 
 \lambda_{H_{1}\phi_{2}} \geq - \lambda_{H_{2}\phi_{2} } \sqrt{\frac{\lambda_{1}}{\lambda_{2}}}  \right)
 \nonumber \\
&& \bigcup 
\biggl( \lambda_{\phi_{2}}, \lambda_{1}, \lambda_{2} > 0; 
2 \sqrt{\lambda_{\phi_{2}} \lambda_{2}} \geq \lambda_{H_{2}\phi_{2}} > -2 \sqrt{\lambda_{\phi_{2}} \lambda_{2}}; - \lambda_{H_{2}\phi_{2}} \sqrt{\frac{\lambda_{1}}{\lambda_{2}}} \geq
 \lambda_{H_{1}\phi_{2}} > - 2 \sqrt{\lambda_{\phi_{2}} \lambda_{1}}; \nonumber \\
&& 2 \lambda_{\phi_{2}} \lambda_{12} > \lambda_{H_{1}\phi_{2}} \lambda_{H_{2}\phi_{2}}
- \sqrt{ \left( \lambda^{2}_{H_{1}\phi_{2}} - 4 \lambda_{\phi_{2}} \lambda_{1} \right)
\left( \lambda^{2}_{H_{2}\phi_{2}} - 4 \lambda_{\phi_{2}} \lambda_{2} \right) }\biggr)\,, 
    \end{eqnarray}

where $i = 1,2$.

{\bf Case II: $\lambda^{\prime}_{12} < 0$}

For this case, the condition will same as before but we need to replace $\label{12}$ by $\lambda_{12}+\lambda^{\prime}_{12}$ {\it i.e.} $\Omega_{2} = \Omega_{1} \bigg|_{\lambda_{12}
\rightarrow
\lambda_{12} + \lambda^{\prime}_{12}}$.

\item {\bf Bound from pertubativity:}

The quartic couplings can be written in terms of the scalar masses and mixing angles in the following 
manner,

\begin{eqnarray}
    \lambda_{1} &=& \frac{1}{2 v^{2}_{1}} \biggl( \frac{\mu v_{2} v_{\phi_{1}}}{\sqrt{2} v_{1}} 
    + \sum^{3}_{i=1} R^{2}_{i1} M^2_{h_{i}}  \biggr) \nonumber \\
    \lambda_{2} &=& \frac{1}{2 v^{2}_{2}} \biggl( \frac{\mu v_{1} v_{\phi_{1}}}{\sqrt{2} v_{2}} 
    + \sum^{3}_{i=1} R^{2}_{i2} M^2_{h_{i}}  \biggr) \nonumber \\
    \lambda_{\phi_{2}} &=& \frac{1}{2 v^{2}_{\phi_{2}}} \sum^{3}_{i=1} R^{2}_{i3} M^2_{h_{i}}
    \nonumber \\
    \lambda_{12} &=& \frac{2 M^2_{H^{\pm}}}{v^{2}} + \frac{\mu v_{\phi_{1}}}{\sqrt{2} v_{1} v_{2}}
    + \frac{1}{v_{1} v_{2} } \sum^{3}_{i=1} R_{i1} R_{i2} M^2_{h_{i}} \nonumber \\
    \lambda^{\prime}_{12} &=& -\frac{2 M^2_{H^{\pm}}}{v^{2}} + \frac{2 M^2_{A}}
    {\left( v^{2} + \frac{v^2_{1} v^2_{2}}{v^2_{\phi_{1}}} \right)} \nonumber \\
    \lambda_{H_{1} \phi_{2} } &=& \frac{1}{v_{1} v_{\phi_{2}}} \sum^{3}_{i=1} R_{i1} R_{i3} M^2_{h_{i}}
    \nonumber \\
    \lambda_{H_{2} \phi_{2} } &=& \frac{1}{v_{2} v_{\phi_{2}}} \sum^{3}_{i=1} R_{i2} R_{i3} M^2_{h_{i}}
    \nonumber \\
\end{eqnarray}

    \end{itemize}

\section{Oblique parameters S, T, and U}

By following Ref. \cite{Grimus:2008nb}, we can write down the scalars in terms of the mass 
eigenstates when the first component in the LHS column matrix matches with the 
SM Higgs neutral component as,

\begin{eqnarray}
    \begin{pmatrix}
        \phi_{a} + i G \\
        \phi_{b} + i A \\
        \phi_{c} + i a \\
        \phi_{d}
    \end{pmatrix}
    = \begin{pmatrix}
        i & V_{11} & V_{12} & V_{13} & 0 & U_{13} & 0 \\
        0 & V_{21} & V_{22} & V_{23} & i & U_{23} & 0 \\
        0 & V_{31} & V_{32} & V_{33} & 0 & U_{33} & i \\
        0 & R_{13} & R_{23} & R_{33} & 0 & 0 & 0
    \end{pmatrix}
    \begin{pmatrix}
        G\\
        h_{1} \\
        h_{2} \\
        h_{3} \\
        A \\
        \phi_{1} \\
        a
    \end{pmatrix}
\end{eqnarray}
where \begin{eqnarray}
&&    V_{11} = U_{11}R_{11} + U_{12}R_{12}, V_{12} = U_{11}R_{21}+U_{12}R_{22}, V_{13} = U_{11}R_{31}+U_{12}R_{32}\,, \nonumber \\ &&
V_{21} = U_{21}R_{11} + U_{22}R_{12}, V_{22} = U_{21}R_{21}+U_{22}R_{22}, V_{23} = U_{21}R_{31}+U_{22}R_{32}\,, \nonumber \\ &&
V_{31} = U_{31}R_{11} + U_{32}R_{12}, V_{32} = U_{31}R_{21}+U_{32}R_{22}, V_{33} = U_{31}R_{31}+U_{32}R_{32}\,.  
\end{eqnarray}
In the above relations, $R_{ij}$ and $U_{ij}$ are matrix which diagonalise the CP-even and CP-odd
mass matrices s shown in Eq.\,(\ref{CP-even-mass}, \ref{cp-odd-matrix}). 
The oblique parameters S, T and U can be expressed as,

\begin{eqnarray}
S &=& \frac{1}{24\pi} \biggl[ (2 s^2_{w} -1)^{2} G\left( M^{2}_{H^{\pm}},M^{2}_{H^{\pm}},M^{2}_{Z} \right) 
+ \biggl( V^2_{21} G\left( M^{2}_{h_{1}},M^{2}_{A},M^{2}_{Z}\right)  
+ V^2_{22} G\left( M^{2}_{h_{2}},M^{2}_{A},M^{2}_{Z} \right) \nonumber \\
&& + V^2_{23} G\left( M^{2}_{h_{3}},M^{2}_{A},M^{2}_{Z} \right)
+ U^2_{23} G\left(M^{2}_{A},M^{2}_{\phi_{1}},M^{2}_{Z} \right) 
\biggr) - 2 \ln{M^{2}_{H^{\pm}}}
+ \biggl(  \left( V^{2}_{11} + V^{2}_{21} \right) \ln{M^2_{h_{1}}} \nonumber \\
&& + \left( V^{2}_{12} + V^{2}_{22} \right) \ln{M^2_{h_{2}}}
+ \left( V^{2}_{13} + V^{2}_{23} \right) \ln{M^2_{h_{3}}} + \ln{M^2_{A}}
+ \left( U^{2}_{13} + U^{2}_{23} \right) \ln{M^2_{\phi_{1}}}
\biggr) - \ln{M^2_{h_{SM}}} \nonumber \\
&& \biggl( V^{2}_{11} \Hat{G}\left(M^2_{h_1},M^2_{Z} \right) 
+ V^{2}_{12} \Hat{G}\left(M^2_{h_2},M^2_{Z} \right)
+ V^{2}_{13} \Hat{G}\left(M^2_{h_3},M^2_{Z} \right)
+ U^{2}_{13} \Hat{G}\left(M^2_{\phi_{1}},M^2_{Z} \right) 
 - \Hat{G}\left( M^2_{h_{SM}}, M^2_{Z} \right)
\biggr) 
\biggr] \nonumber \\
\end{eqnarray}

\begin{eqnarray}
    T &=& \frac{1}{16 \pi^{2} M^2_{W} s^2_{W}} \biggl[ \biggl(
    V^2_{21} F\left( M^2_{H^{\pm}}, M^2_{h_{1}} \right) 
    + V^2_{22} F\left( M^2_{H^{\pm}}, M^2_{h_{2}} \right)
    + V^2_{23} F\left( M^2_{H^{\pm}}, M^2_{h_{3}} \right)
    + F\left( M^2_{H^{\pm}}, M^2_{A} \right) \nonumber \\
   && + U^2_{23} F\left( M^2_{H^{\pm}}, M^2_{\phi_{1}} \right) \biggr)
   - \biggl(
 V^2_{21} F\left( M^2_{h_{1}}, M^2_{A} \right)
 + V^2_{22} F\left( M^2_{h_{2}}, M^2_{A} \right)
 + V^2_{23} F\left( M^2_{h_{3}}, M^2_{A} \right) \nonumber \\
&& + U^2_{23} F\left(M^2_{A}, M^2_{\phi_{1}} \right)
   \biggr) + 3 \biggl( V^2_{11} \left( F(M^2_{Z},M^2_{h_{1}}) - F(M^2_{W},M^2_{h_{1}}) \right)
   + V^2_{12} \left( F(M^2_{Z},M^2_{h_{2}}) - F(M^2_{W},M^2_{h_{2}}) \right) \nonumber \\
  && + V^2_{13} \left( F(M^2_{Z},M^2_{h_{3}}) - F(M^2_{W},M^2_{h_{3}}) \right)
  + U^2_{13} \left( F(M^2_{Z},M^2_{\phi_{1}}) - F(M^2_{W},M^2_{\phi_{1}}) \right) \nonumber \\
 &&  -  \left( F(M^2_{Z},M^2_{h_{SM}}) - F(M^2_{W},M^2_{h_{SM}}) \right)
   \biggr)
    \biggr]
\end{eqnarray}

\begin{eqnarray}
    U &=& \frac{1}{24 \pi} \biggl[ \biggl(
    V^{2}_{21} G\left( M^2_{H^{\pm}}, M^2_{h_{1}}, M^2_{W} \right)
    + V^{2}_{22} G\left( M^2_{H^{\pm}}, M^2_{h_{2}}, M^2_{W} \right)
    + V^{2}_{23} G\left( M^2_{H^{\pm}}, M^2_{h_{3}}, M^2_{W} \right) \nonumber \\
   && + G\left( M^2_{H^{\pm}}, M^2_{A}, M^2_{W} \right) 
   + U^{2}_{23} G\left( M^2_{H^{\pm}}, M^2_{\phi_{1}}, M^2_{W} \right)
   - (2 s^2_{W} -1)^{2} G\left( M^2_{H^{\pm}}, M^2_{H^{\pm}}, M^2_{Z} \right) \biggr) \nonumber \\
&&  - \biggl(
   V^2_{21} G\left( M^2_{h_{1}}, M^2_{A}, M^2_{Z} \right)
   +    V^2_{22} G\left( M^2_{h_{2}}, M^2_{A}, M^2_{Z} \right)
   +    V^2_{23} G\left( M^2_{h_{3}}, M^2_{A}, M^2_{Z} \right)
   +    U^2_{23} G\left( M^2_{A}, M^2_{\phi_{1}}, M^2_{Z} \right)
   \biggr) \nonumber \\
   && + \biggl(
V^2_{11} \left( \Hat{G}(M^2_{h_1},M^2_{W}) -\Hat{G}(M^2_{h_1},M^2_{Z})  \right)
+ V^2_{12} \left( \Hat{G}(M^2_{h_2},M^2_{W}) -\Hat{G}(M^2_{h_2},M^2_{Z})  \right)\nonumber \\
&&
+ V^2_{13} \left( \Hat{G}(M^2_{h_3},M^2_{W}) -\Hat{G}(M^2_{h_3},M^2_{Z})  \right)  + U^2_{13} \left( \Hat{G}(M^2_{\phi_1},M^2_{W}) -\Hat{G}(M^2_{\phi_1},M^2_{Z})  \right) \nonumber \\ &&
-\left( \Hat{G}(M^2_{h_{SM}},M^2_{W}) -\Hat{G}(M^2_{h_{SM}},M^2_{Z})  \right)
   \biggr)
       \biggr]
\end{eqnarray}

The explicit form of the functions $F(x,y)$, $G(x,y,z)$ and $\Hat{G}(x,z)$ are given by,
\begin{eqnarray}
    F(x,y)&=& 
\begin{cases}
    \frac{x+y}{2} - \frac{xy}{x-y} \ln{\frac{x}{y}},& \text{for } x\neq y\\
    0,              & \text{for } x = y
\end{cases}
\nonumber \\
 G(x,y,z)&=& 
\begin{cases}
    -\frac{16}{3} + 5 \frac{x+y}{z} - 2 \frac{(x-y)^2}{z^{2}} + \frac{3}{z} \biggl[
    \frac{x^{2}+y^{2}}{x-y} - \frac{x^{2} - y^{2}}{z} + \frac{(x-y)^{3}}{3 z^{2}}
    \biggr] \ln{\frac{x}{y}} + \frac{r}{z^{3}} f(p,q),& \text{for } x\neq y\\
    -\frac{16}{3} + \frac{16}{z}x + \frac{r}{z^{3}} f(p,q) ,              & \text{for } x = y
\end{cases}
\nonumber \\
\Hat{G}(x,z) &=& -\frac{79}{3} + 9 \frac{x}{z} - 2 \frac{x^{2}}{z^{2}} 
+ \left( -10 + 18 \frac{x}{z} - 6 \frac{x^{2}}{z^{2}} + \frac{x^{3}}{z^{3}} 
- 9 \frac{x+z}{x-z} \right) \ln{\frac{x}{z}} \nonumber \\
&& + \left( 12 - 4 \frac{x}{z} + \frac{x^{2}}{z^{2}} \right) \frac{f(x,x^{2} - 4 x z)}{z}
\end{eqnarray}

where $p = x+y-z$, $q = z^{2} - 2 z (x+y) + (x-y)^{2}$ and the function 
$f(p,q)$ can be expressed as,
\[
    f(p,q)= 
\begin{cases}
    \sqrt{q} \ln{|\frac{p-\sqrt{q}}{p+\sqrt{q}}|},& \text{for } q\neq 0\\
    0,              & \text{for } q = 0 \\
    2 \sqrt{-q}\, tan^{-1} \frac{\sqrt{-q}}{p}, & \text{for } q < 0 \,.
\end{cases}
\]


\begin{thebibliography}{99}

\bibitem{Bertone:2016nfn}
G.~Bertone and D.~Hooper,
Rev. Mod. Phys. \textbf{90}, no.4, 045002 (2018)
doi:10.1103/RevModPhys.90.045002
[arXiv:1605.04909 [astro-ph.CO]].

\bibitem{Kajita:2016cak}
T.~Kajita,
Rev. Mod. Phys. \textbf{88}, no.3, 030501 (2016)
doi:10.1103/RevModPhys.88.030501

\bibitem{McDonald:2016ixn}
A.~B.~McDonald,
Rev. Mod. Phys. \textbf{88}, no.3, 030502 (2016)
doi:10.1103/RevModPhys.88.030502

\bibitem{Esteban:2020cvm}
I.~Esteban, M.~C.~Gonzalez-Garcia, M.~Maltoni, T.~Schwetz and A.~Zhou,
JHEP \textbf{09}, 178 (2020)
doi:10.1007/JHEP09(2020)178
[arXiv:2007.14792 [hep-ph]].

\bibitem{Aoyama:2020ynm}
T.~Aoyama, N.~Asmussen, M.~Benayoun, J.~Bijnens, T.~Blum, M.~Bruno, I.~Caprini, C.~M.~Carloni Calame, M.~C\`e and G.~Colangelo, \textit{et al.}
Phys. Rept. \textbf{887}, 1-166 (2020)
doi:10.1016/j.physrep.2020.07.006
[arXiv:2006.04822 [hep-ph]].

\bibitem{CDF:2022hxs}
T.~Aaltonen \textit{et al.} [CDF],
Science \textbf{376}, no.6589, 170-176 (2022)
doi:10.1126/science.abk1781

\bibitem{Sofue:2000jx}
Y.~Sofue and V.~Rubin,
Ann. Rev. Astron. Astrophys. \textbf{39}, 137-174 (2001)
doi:10.1146/annurev.astro.39.1.137
[arXiv:astro-ph/0010594 [astro-ph]].

\bibitem{Clowe:2003tk}
D.~Clowe, A.~Gonzalez and M.~Markevitch,
Astrophys. J. \textbf{604}, 596-603 (2004)
doi:10.1086/381970
[arXiv:astro-ph/0312273 [astro-ph]].

\bibitem{Planck:2018vyg}
N.~Aghanim \textit{et al.} [Planck],
Astron. Astrophys. \textbf{641}, A6 (2020)
[erratum: Astron. Astrophys. \textbf{652}, C4 (2021)]
doi:10.1051/0004-6361/201833910
[arXiv:1807.06209 [astro-ph.CO]].

\bibitem{Baker:2006ts}
C.~A.~Baker, D.~D.~Doyle, P.~Geltenbort, K.~Green, M.~G.~D.~van der Grinten, P.~G.~Harris, P.~Iaydjiev, S.~N.~Ivanov, D.~J.~R.~May and J.~M.~Pendlebury, \textit{et al.}
Phys. Rev. Lett. \textbf{97}, 131801 (2006)
doi:10.1103/PhysRevLett.97.131801
[arXiv:hep-ex/0602020 [hep-ex]].

\bibitem{Super-Kamiokande:1998kpq}
Y.~Fukuda \textit{et al.} [Super-Kamiokande],
Phys. Rev. Lett. \textbf{81}, 1562-1567 (1998)
doi:10.1103/PhysRevLett.81.1562
[arXiv:hep-ex/9807003 [hep-ex]].

\bibitem{SNO:2002tuh}
Q.~R.~Ahmad \textit{et al.} [SNO],
Phys. Rev. Lett. \textbf{89}, 011301 (2002)
doi:10.1103/PhysRevLett.89.011301
[arXiv:nucl-ex/0204008 [nucl-ex]].

\bibitem{KamLAND:2002uet}
K.~Eguchi \textit{et al.} [KamLAND],
Phys. Rev. Lett. \textbf{90}, 021802 (2003)
doi:10.1103/PhysRevLett.90.021802
[arXiv:hep-ex/0212021 [hep-ex]].

\bibitem{DayaBay:2015lja}
F.~P.~An \textit{et al.} [Daya Bay],
Phys. Rev. Lett. \textbf{116}, no.6, 061801 (2016)
[erratum: Phys. Rev. Lett. \textbf{118}, no.9, 099902 (2017)]
doi:10.1103/PhysRevLett.116.061801
[arXiv:1508.04233 [hep-ex]].

\bibitem{Muong-2:2021ojo}
B.~Abi \textit{et al.} [Muon g-2],
Phys. Rev. Lett. \textbf{126}, no.14, 141801 (2021)
doi:10.1103/PhysRevLett.126.141801
[arXiv:2104.03281 [hep-ex]].

\bibitem{Borsanyi:2020mff}
S.~Borsanyi, Z.~Fodor, J.~N.~Guenther, C.~Hoelbling, S.~D.~Katz, L.~Lellouch, T.~Lippert, K.~Miura, L.~Parato and K.~K.~Szabo, \textit{et al.}
Nature \textbf{593}, no.7857, 51-55 (2021)
doi:10.1038/s41586-021-03418-1
[arXiv:2002.12347 [hep-lat]].

\bibitem{CMD-3:2023alj}
F.~V.~Ignatov \textit{et al.} [CMD-3],
[arXiv:2302.08834 [hep-ex]].

\bibitem{Peccei:1977hh}
R.~D.~Peccei and H.~R.~Quinn,
Phys. Rev. Lett. \textbf{38}, 1440-1443 (1977)
doi:10.1103/PhysRevLett.38.1440

\bibitem{Peccei:1977ur}
R.~D.~Peccei and H.~R.~Quinn,
Phys. Rev. D \textbf{16}, 1791-1797 (1977)
doi:10.1103/PhysRevD.16.1791

\bibitem{Weinberg:1977ma}
S.~Weinberg,
Phys. Rev. Lett. \textbf{40}, 223-226 (1978)
doi:10.1103/PhysRevLett.40.223

\bibitem{Wilczek:1977pj}
F.~Wilczek,
Phys. Rev. Lett. \textbf{40}, 279-282 (1978)
doi:10.1103/PhysRevLett.40.279

\bibitem{Glashow:1976nt}
S.~L.~Glashow and S.~Weinberg,
Phys. Rev. D \textbf{15}, 1958 (1977)
doi:10.1103/PhysRevD.15.1958

\bibitem{Paschos:1976ay}
E.~A.~Paschos,
Phys. Rev. D \textbf{15}, 1966 (1977)
doi:10.1103/PhysRevD.15.1966

\bibitem{Bjorkeroth:2017tsz}
F.~Bj\"orkeroth, E.~J.~Chun and S.~F.~King,
Phys. Lett. B \textbf{777}, 428-434 (2018)
doi:10.1016/j.physletb.2017.12.058
[arXiv:1711.05741 [hep-ph]].

\bibitem{Harigaya:2013vja}
K.~Harigaya, M.~Ibe, K.~Schmitz and T.~T.~Yanagida,
Phys. Rev. D \textbf{88}, no.7, 075022 (2013)
doi:10.1103/PhysRevD.88.075022
[arXiv:1308.1227 [hep-ph]].

\bibitem{Dias:2002hz}
A.~G.~Dias, V.~Pleitez and M.~D.~Tonasse,
Phys. Rev. D \textbf{69}, 015007 (2004)
doi:10.1103/PhysRevD.69.015007
[arXiv:hep-ph/0210172 [hep-ph]].

\bibitem{Ahn:2014gva}
Y.~H.~Ahn,
Phys. Rev. D \textbf{91}, 056005 (2015)
doi:10.1103/PhysRevD.91.056005
[arXiv:1410.1634 [hep-ph]].

\bibitem{Honecker:2015ela}
G.~Honecker and W.~Staessens,
J. Phys. Conf. Ser. \textbf{631}, no.1, 012080 (2015)
doi:10.1088/1742-6596/631/1/012080
[arXiv:1502.00985 [hep-th]].

\bibitem{Clarke:2015bea}
J.~D.~Clarke and R.~R.~Volkas,
Phys. Rev. D \textbf{93}, no.3, 035001 (2016)
doi:10.1103/PhysRevD.93.035001
[arXiv:1509.07243 [hep-ph]].

\bibitem{Hawking:1987mz}
S.~W.~Hawking,
Phys. Lett. B \textbf{195}, 337 (1987)
doi:10.1016/0370-2693(87)90028-1

\bibitem{Lavrelashvili:1987jg}
G.~V.~Lavrelashvili, V.~A.~Rubakov and P.~G.~Tinyakov,
JETP Lett. \textbf{46}, 167-169 (1987)

\bibitem{Giddings:1988cx}
S.~B.~Giddings and A.~Strominger,
Nucl. Phys. B \textbf{307}, 854-866 (1988)
doi:10.1016/0550-3213(88)90109-5

\bibitem{Coleman:1988tj}
S.~R.~Coleman,
Nucl. Phys. B \textbf{310}, 643-668 (1988)
doi:10.1016/0550-3213(88)90097-1

\bibitem{Gilbert:1989nq}
G.~Gilbert,
Nucl. Phys. B \textbf{328}, 159-170 (1989)
doi:10.1016/0550-3213(89)90097-7

\bibitem{Dias:2021lmf}
A.~G.~Dias, J.~Leite and D.~S.~V.~Gon\c{c}alves,
Phys. Rev. D \textbf{104}, no.7, 075014 (2021)
doi:10.1103/PhysRevD.104.075014
[arXiv:2106.07518 [hep-ph]].

\bibitem{HFLAV:2016hnz}
Y.~Amhis \textit{et al.} [HFLAV],
Eur. Phys. J. C \textbf{77}, no.12, 895 (2017)
doi:10.1140/epjc/s10052-017-5058-4
[arXiv:1612.07233 [hep-ex]].

\bibitem{Misiak:2017bgg}
M.~Misiak and M.~Steinhauser,
Eur. Phys. J. C \textbf{77}, no.3, 201 (2017)
doi:10.1140/epjc/s10052-017-4776-y
[arXiv:1702.04571 [hep-ph]].

\bibitem{Zhitnitsky:1980tq}
A.~R.~Zhitnitsky,
Sov. J. Nucl. Phys. \textbf{31}, 260 (1980)

\bibitem{Dine:1981rt}
M.~Dine, W.~Fischler and M.~Srednicki,
Phys. Lett. B \textbf{104}, 199-202 (1981)
doi:10.1016/0370-2693(81)90590-6

\bibitem{Chang:2018rso}
J.~H.~Chang, R.~Essig and S.~D.~McDermott,
JHEP \textbf{09}, 051 (2018)
doi:10.1007/JHEP09(2018)051
[arXiv:1803.00993 [hep-ph]].

\bibitem{Branco:2011iw}
G.~C.~Branco, P.~M.~Ferreira, L.~Lavoura, M.~N.~Rebelo, M.~Sher and J.~P.~Silva,
Phys. Rept. \textbf{516}, 1-102 (2012)
doi:10.1016/j.physrep.2012.02.002
[arXiv:1106.0034 [hep-ph]].

\bibitem{Preskill:1982cy}
J.~Preskill, M.~B.~Wise and F.~Wilczek,
Phys. Lett. B \textbf{120}, 127-132 (1983)
doi:10.1016/0370-2693(83)90637-8

\bibitem{Abbott:1982af}
L.~F.~Abbott and P.~Sikivie,
Phys. Lett. B \textbf{120}, 133-136 (1983)
doi:10.1016/0370-2693(83)90638-X

\bibitem{DiLuzio:2020wdo}
L.~Di Luzio, M.~Giannotti, E.~Nardi and L.~Visinelli,
Phys. Rept. \textbf{870}, 1-117 (2020)
doi:10.1016/j.physrep.2020.06.002
[arXiv:2003.01100 [hep-ph]].

\bibitem{Marsh:2015xka}
D.~J.~E.~Marsh,
Phys. Rept. \textbf{643}, 1-79 (2016)
doi:10.1016/j.physrep.2016.06.005
[arXiv:1510.07633 [astro-ph.CO]].

\bibitem{Muhlleitner:2016mzt}
M.~Muhlleitner, M.~O.~P.~Sampaio, R.~Santos and J.~Wittbrodt,
JHEP \textbf{03}, 094 (2017)
doi:10.1007/JHEP03(2017)094
[arXiv:1612.01309 [hep-ph]].

\bibitem{Arhrib:2018qmw}
A.~Arhrib, R.~Benbrik, M.~El Kacimi, L.~Rahili and S.~Semlali,
Eur. Phys. J. C \textbf{80}, no.1, 13 (2020)
doi:10.1140/epjc/s10052-019-7472-2
[arXiv:1811.12431 [hep-ph]].

\bibitem{Maki:1962mu}
Z.~Maki, M.~Nakagawa and S.~Sakata,
Prog. Theor. Phys. \textbf{28}, 870-880 (1962)
doi:10.1143/PTP.28.870

\bibitem{Biswas:2016yan}
A.~Biswas, S.~Choubey and S.~Khan,
JHEP \textbf{09}, 147 (2016)
doi:10.1007/JHEP09(2016)147
[arXiv:1608.04194 [hep-ph]].

\bibitem{Deppisch:2015qwa}
F.~F.~Deppisch, P.~S.~Bhupal Dev and A.~Pilaftsis,
New J. Phys. \textbf{17}, no.7, 075019 (2015)
doi:10.1088/1367-2630/17/7/075019
[arXiv:1502.06541 [hep-ph]].

\bibitem{Gorbunov:2007ak}
D.~Gorbunov and M.~Shaposhnikov,
JHEP \textbf{10}, 015 (2007)
[erratum: JHEP \textbf{11}, 101 (2013)]
doi:10.1088/1126-6708/2007/10/015
[arXiv:0705.1729 [hep-ph]].

\bibitem{Boyarsky:2009ix}
A.~Boyarsky, O.~Ruchayskiy and M.~Shaposhnikov,
Ann. Rev. Nucl. Part. Sci. \textbf{59}, 191-214 (2009)
doi:10.1146/annurev.nucl.010909.083654
[arXiv:0901.0011 [hep-ph]].

\bibitem{Ruchayskiy:2012si}
O.~Ruchayskiy and A.~Ivashko,
JCAP \textbf{10}, 014 (2012)
doi:10.1088/1475-7516/2012/10/014
[arXiv:1202.2841 [hep-ph]].

\bibitem{Domcke:2020ety}
V.~Domcke, M.~Drewes, M.~Hufnagel and M.~Lucente,
JHEP \textbf{01}, 200 (2021)
doi:10.1007/JHEP01(2021)200
[arXiv:2009.11678 [hep-ph]].

\bibitem{Foot:2013hna}
R.~Foot, A.~Kobakhidze, K.~L.~McDonald and R.~R.~Volkas,
Phys. Rev. D \textbf{89}, no.11, 115018 (2014)
doi:10.1103/PhysRevD.89.115018
[arXiv:1310.0223 [hep-ph]].

\bibitem{Kim:1979if}
J.~E.~Kim,
Phys. Rev. Lett. \textbf{43}, 103 (1979)
doi:10.1103/PhysRevLett.43.103

\bibitem{Shifman:1979if}
M.~A.~Shifman, A.~I.~Vainshtein and V.~I.~Zakharov,
Nucl. Phys. B \textbf{166}, 493-506 (1980)
doi:10.1016/0550-3213(80)90209-6

\bibitem{Covi:2022hqb}
L.~Covi and S.~Khan,
JCAP \textbf{09}, 064 (2022)
doi:10.1088/1475-7516/2022/09/064
[arXiv:2205.10150 [hep-ph]].

\bibitem{Turner:1985si}
M.~S.~Turner,
Phys. Rev. D \textbf{33}, 889-896 (1986)
doi:10.1103/PhysRevD.33.889

\bibitem{GrillidiCortona:2015jxo}
G.~Grilli di Cortona, E.~Hardy, J.~Pardo Vega and G.~Villadoro,
JHEP \textbf{01}, 034 (2016)
doi:10.1007/JHEP01(2016)034
[arXiv:1511.02867 [hep-ph]].

\bibitem{Gorghetto:2018myk}
M.~Gorghetto, E.~Hardy and G.~Villadoro,
JHEP \textbf{07}, 151 (2018)
doi:10.1007/JHEP07(2018)151
[arXiv:1806.04677 [hep-ph]].

\bibitem{Buschmann:2019icd}
M.~Buschmann, J.~W.~Foster and B.~R.~Safdi,
Phys. Rev. Lett. \textbf{124}, no.16, 161103 (2020)
doi:10.1103/PhysRevLett.124.161103
[arXiv:1906.00967 [astro-ph.CO]].

\bibitem{Gorghetto:2020qws}
M.~Gorghetto, E.~Hardy and G.~Villadoro,
SciPost Phys. \textbf{10}, no.2, 050 (2021)
doi:10.21468/SciPostPhys.10.2.050
[arXiv:2007.04990 [hep-ph]].

\bibitem{Buschmann:2021sdq}
M.~Buschmann, J.~W.~Foster, A.~Hook, A.~Peterson, D.~E.~Willcox, W.~Zhang and B.~R.~Safdi,
Nature Commun. \textbf{13}, no.1, 1049 (2022)
doi:10.1038/s41467-022-28669-y
[arXiv:2108.05368 [hep-ph]].



\bibitem{Kawasaki:2013ae}
M.~Kawasaki and K.~Nakayama,
Ann. Rev. Nucl. Part. Sci. \textbf{63}, 69-95 (2013)
doi:10.1146/annurev-nucl-102212-170536
[arXiv:1301.1123 [hep-ph]].

\bibitem{Kawasaki:2018qwp}
M.~Kawasaki, E.~Sonomoto and T.~T.~Yanagida,
Phys. Lett. B \textbf{782}, 181-184 (2018)
doi:10.1016/j.physletb.2018.05.014
[arXiv:1801.07409 [hep-ph]].

\bibitem{Planck:2015sxf}
P.~A.~R.~Ade \textit{et al.} [Planck],
Astron. Astrophys. \textbf{594}, A20 (2016)
doi:10.1051/0004-6361/201525898
[arXiv:1502.02114 [astro-ph.CO]].

\bibitem{Planck:2018jri}
Y.~Akrami \textit{et al.} [Planck],
Astron. Astrophys. \textbf{641}, A10 (2020)
doi:10.1051/0004-6361/201833887
[arXiv:1807.06211 [astro-ph.CO]].

\bibitem{Planck:2015fie}
P.~A.~R.~Ade \textit{et al.} [Planck],
Astron. Astrophys. \textbf{594}, A13 (2016)
doi:10.1051/0004-6361/201525830
[arXiv:1502.01589 [astro-ph.CO]].

\bibitem{ADMX:2018gho}
N.~Du \textit{et al.} [ADMX],
Phys. Rev. Lett. \textbf{120}, no.15, 151301 (2018)
doi:10.1103/PhysRevLett.120.151301
[arXiv:1804.05750 [hep-ex]].

\bibitem{ADMX:2019uok}
T.~Braine \textit{et al.} [ADMX],
Phys. Rev. Lett. \textbf{124}, no.10, 101303 (2020)
doi:10.1103/PhysRevLett.124.101303
[arXiv:1910.08638 [hep-ex]].

\bibitem{ADMX:2021nhd}
C.~Bartram \textit{et al.} [ADMX],
Phys. Rev. Lett. \textbf{127}, no.26, 261803 (2021)
doi:10.1103/PhysRevLett.127.261803
[arXiv:2110.06096 [hep-ex]].

\bibitem{Lee:2019mfy}
S.~Lee, S.~Ahn, J.~Choi, B.~Rok Ko and Y.~K.~Semertzidis,
PoS \textbf{EPS-HEP2019}, 101 (2020)
doi:10.22323/1.364.0101
[arXiv:1910.00047 [hep-ex]].

\bibitem{Semertzidis:2019gkj}
Y.~K.~Semertzidis, J.~E.~Kim, S.~Youn, J.~Choi, W.~Chung, S.~Haciomeroglu, D.~Kim, J.~Kim, B.~Ko and O.~Kwon, \textit{et al.}
[arXiv:1910.11591 [physics.ins-det]].

\bibitem{MADMAX:2019pub}
P.~Brun \textit{et al.} [MADMAX],
Eur. Phys. J. C \textbf{79}, no.3, 186 (2019)
doi:10.1140/epjc/s10052-019-6683-x
[arXiv:1901.07401 [physics.ins-det]].

\bibitem{Alesini:2019nzq}
D.~Alesini, D.~Babusci, P.~Beltrame, S.J., F.~Bj\"orkeroth, F.~Bossi, P.~Ciambrone, G.~Delle Monache, D.~Di Gioacchino, P.~Falferi and A.~Gallo, \textit{et al.}
[arXiv:1911.02427 [physics.ins-det]].

\bibitem{Ouellet:2018beu}
J.~L.~Ouellet, C.~P.~Salemi, J.~W.~Foster, R.~Henning, Z.~Bogorad, J.~M.~Conrad, J.~A.~Formaggio, Y.~Kahn, J.~Minervini and A.~Radovinsky, \textit{et al.}
Phys. Rev. Lett. \textbf{122}, no.12, 121802 (2019)
doi:10.1103/PhysRevLett.122.121802
[arXiv:1810.12257 [hep-ex]].

\bibitem{Charpak:1962zz}
G.~Charpak, F.~J.~M.~Farley and R.~L.~Garwin,
Phys. Lett. \textbf{1}, 16 (1962)
doi:10.1016/0031-9163(62)90263-9

\bibitem{Charpak:1965zz}
G.~Charpak, P.~J.~M.~Farley, E.~L.~Garwin, T.~Muller, J.~C.~Sens and A.~Zichichi,
Nuovo Cim. \textbf{37}, 1241-1363 (1965)
doi:10.1007/BF02783344

\bibitem{Combley:1974tw}
F.~Combley and E.~Picasso,
Phys. Rept. \textbf{14}, 1 (1974)
doi:10.1016/0370-1573(74)90004-0

\bibitem{CERN-Mainz-Daresbury:1978ccd}
J.~Bailey \textit{et al.} [CERN-Mainz-Daresbury],
Nucl. Phys. B \textbf{150}, 1-75 (1979)
doi:10.1016/0550-3213(79)90292-X

\bibitem{Muong-2:2006rrc}
G.~W.~Bennett \textit{et al.} [Muon g-2],
Phys. Rev. D \textbf{73}, 072003 (2006)
doi:10.1103/PhysRevD.73.072003
[arXiv:hep-ex/0602035 [hep-ex]].

\bibitem{Davier:2017zfy}
M.~Davier, A.~Hoecker, B.~Malaescu and Z.~Zhang,
Eur. Phys. J. C \textbf{77}, no.12, 827 (2017)
doi:10.1140/epjc/s10052-017-5161-6
[arXiv:1706.09436 [hep-ph]].

\bibitem{Keshavarzi:2018mgv}
A.~Keshavarzi, D.~Nomura and T.~Teubner,
Phys. Rev. D \textbf{97}, no.11, 114025 (2018)
doi:10.1103/PhysRevD.97.114025
[arXiv:1802.02995 [hep-ph]].

\bibitem{Colangelo:2018mtw}
G.~Colangelo, M.~Hoferichter and P.~Stoffer,
JHEP \textbf{02}, 006 (2019)
doi:10.1007/JHEP02(2019)006
[arXiv:1810.00007 [hep-ph]].

\bibitem{Davier:2019can}
M.~Davier, A.~Hoecker, B.~Malaescu and Z.~Zhang,
Eur. Phys. J. C \textbf{80}, no.3, 241 (2020)
[erratum: Eur. Phys. J. C \textbf{80}, no.5, 410 (2020)]
doi:10.1140/epjc/s10052-020-7792-2
[arXiv:1908.00921 [hep-ph]].

\bibitem{Arcadi:2022lpp}
G.~Arcadi, N.~Benincasa, A.~Djouadi and K.~Kannike,
Phys. Rev. D \textbf{108}, no.5, 055010 (2023)
doi:10.1103/PhysRevD.108.055010
[arXiv:2212.14788 [hep-ph]].

\bibitem{Arcadi:2022dmt}
G.~Arcadi and A.~Djouadi,
Phys. Rev. D \textbf{106}, no.9, 095008 (2022)
doi:10.1103/PhysRevD.106.095008
[arXiv:2204.08406 [hep-ph]].

\bibitem{Arcadi:2021zdk}
G.~Arcadi, A.~Djouadi and F.~d.~Queiroz,
Phys. Lett. B \textbf{834}, 137436 (2022)
doi:10.1016/j.physletb.2022.137436
[arXiv:2112.11902 [hep-ph]].

\bibitem{Arcadi:2021yyr}
G.~Arcadi, \'A.~S.~de Jesus, T.~B.~de Melo, F.~S.~Queiroz and Y.~S.~Villamizar,
Nucl. Phys. B \textbf{982}, 115882 (2022)
doi:10.1016/j.nuclphysb.2022.115882
[arXiv:2104.04456 [hep-ph]].

\bibitem{Biswas:2016yjr}
A.~Biswas, S.~Choubey and S.~Khan,
JHEP \textbf{02}, 123 (2017)
doi:10.1007/JHEP02(2017)123
[arXiv:1612.03067 [hep-ph]].

\bibitem{Biswas:2017ait}
A.~Biswas, S.~Choubey, L.~Covi and S.~Khan,
JCAP \textbf{02}, 002 (2018)
doi:10.1088/1475-7516/2018/02/002
[arXiv:1711.00553 [hep-ph]].

\bibitem{Biswas:2021dan}
A.~Biswas and S.~Khan,
JHEP \textbf{07}, 037 (2022)
doi:10.1007/JHEP07(2022)037
[arXiv:2112.08393 [hep-ph]].

\bibitem{Costa:2022oaa}
F.~Costa, S.~Khan and J.~Kim,
JHEP \textbf{06}, 026 (2022)
doi:10.1007/JHEP06(2022)026
[arXiv:2202.13126 [hep-ph]].

\bibitem{Lautrup:1971jf}
B.~e.~Lautrup, A.~Peterman and E.~de Rafael,
Phys. Rept. \textbf{3}, 193-259 (1972)
doi:10.1016/0370-1573(72)90011-7

\bibitem{Leveille:1977rc}
J.~P.~Leveille,
Nucl. Phys. B \textbf{137}, 63-76 (1978)
doi:10.1016/0550-3213(78)90051-2

\bibitem{Dedes:2001nx}
A.~Dedes and H.~E.~Haber,
JHEP \textbf{05}, 006 (2001)
doi:10.1088/1126-6708/2001/05/006
[arXiv:hep-ph/0102297 [hep-ph]].

\bibitem{Broggio:2014mna}
A.~Broggio, E.~J.~Chun, M.~Passera, K.~M.~Patel and S.~K.~Vempati,
JHEP \textbf{11}, 058 (2014)
doi:10.1007/JHEP11(2014)058
[arXiv:1409.3199 [hep-ph]].

\bibitem{Ilisie:2015tra}
V.~Ilisie,
JHEP \textbf{04}, 077 (2015)
doi:10.1007/JHEP04(2015)077
[arXiv:1502.04199 [hep-ph]].

\bibitem{Chang:2000ii}
D.~Chang, W.~F.~Chang, C.~H.~Chou and W.~Y.~Keung,
Phys. Rev. D \textbf{63}, 091301 (2001)
doi:10.1103/PhysRevD.63.091301
[arXiv:hep-ph/0009292 [hep-ph]].

\bibitem{Cheung:2001hz}
K.~m.~Cheung, C.~H.~Chou and O.~C.~W.~Kong,
Phys. Rev. D \textbf{64}, 111301 (2001)
doi:10.1103/PhysRevD.64.111301
[arXiv:hep-ph/0103183 [hep-ph]].

\bibitem{Cheung:2003pw}
K.~Cheung and O.~C.~W.~Kong,
Phys. Rev. D \textbf{68}, 053003 (2003)
doi:10.1103/PhysRevD.68.053003
[arXiv:hep-ph/0302111 [hep-ph]].

\bibitem{Gninenko:2001hx}
S.~N.~Gninenko and N.~V.~Krasnikov,
Phys. Lett. B \textbf{513}, 119 (2001)
doi:10.1016/S0370-2693(01)00693-1
[arXiv:hep-ph/0102222 [hep-ph]].

\bibitem{Baek:2001kca}
S.~Baek, N.~G.~Deshpande, X.~G.~He and P.~Ko,
Phys. Rev. D \textbf{64}, 055006 (2001)
doi:10.1103/PhysRevD.64.055006
[arXiv:hep-ph/0104141 [hep-ph]].

\bibitem{1503.01789}
M.~Misiak, H.~M.~Asatrian, R.~Boughezal, M.~Czakon, T.~Ewerth, A.~Ferroglia, P.~Fiedler, P.~Gambino, C.~Greub and U.~Haisch, \textit{et al.}
Phys. Rev. Lett. \textbf{114}, no.22, 221801 (2015)
doi:10.1103/PhysRevLett.114.221801
[arXiv:1503.01789 [hep-ph]].

\bibitem{Altmannshofer:2014pba}
W.~Altmannshofer, S.~Gori, M.~Pospelov and I.~Yavin,
Phys. Rev. Lett. \textbf{113}, 091801 (2014)
doi:10.1103/PhysRevLett.113.091801
[arXiv:1406.2332 [hep-ph]].

\bibitem{Gninenko:2014pea}
S.~N.~Gninenko, N.~V.~Krasnikov and V.~A.~Matveev,
Phys. Rev. D \textbf{91}, 095015 (2015)
doi:10.1103/PhysRevD.91.095015
[arXiv:1412.1400 [hep-ph]].

\bibitem{Kahn:2018cqs}
Y.~Kahn, G.~Krnjaic, N.~Tran and A.~Whitbeck,
JHEP \textbf{09}, 153 (2018)
doi:10.1007/JHEP09(2018)153
[arXiv:1804.03144 [hep-ph]].

\bibitem{MuonCollider:2022xlm}
J.~de Blas \textit{et al.} [Muon Collider],
[arXiv:2203.07261 [hep-ph]].

\bibitem{CDF:2013dpa}
T.~A.~Aaltonen \textit{et al.} [CDF and D0],
Phys. Rev. D \textbf{88}, no.5, 052018 (2013)
doi:10.1103/PhysRevD.88.052018
[arXiv:1307.7627 [hep-ex]].

\bibitem{ATLAS:2017rzl}
M.~Aaboud \textit{et al.} [ATLAS],
Eur. Phys. J. C \textbf{78}, no.2, 110 (2018)
[erratum: Eur. Phys. J. C \textbf{78}, no.11, 898 (2018)]
doi:10.1140/epjc/s10052-017-5475-4
[arXiv:1701.07240 [hep-ex]].

\bibitem{LHCb:2021bjt}
R.~Aaij \textit{et al.} [LHCb],
JHEP \textbf{01}, 036 (2022)
doi:10.1007/JHEP01(2022)036
[arXiv:2109.01113 [hep-ex]].

\bibitem{CDF:2013bqv}
T.~A.~Aaltonen \textit{et al.} [CDF],
Phys. Rev. D \textbf{89}, no.7, 072003 (2014)
doi:10.1103/PhysRevD.89.072003
[arXiv:1311.0894 [hep-ex]].

\bibitem{ATLAS:2023fsi}
 [ATLAS],
ATLAS-CONF-2023-004.

\bibitem{Kownacki:2016pmx}
C.~Kownacki, E.~Ma, N.~Pollard and M.~Zakeri,
Phys. Lett. B \textbf{766}, 149-152 (2017)
doi:10.1016/j.physletb.2017.01.013
[arXiv:1611.05017 [hep-ph]].

\bibitem{Grimus:2008nb}
W.~Grimus, L.~Lavoura, O.~M.~Ogreid and P.~Osland,
Nucl. Phys. B \textbf{801}, 81-96 (2008)
doi:10.1016/j.nuclphysb.2008.04.019
[arXiv:0802.4353 [hep-ph]].

\bibitem{Asadi:2022xiy}
P.~Asadi, C.~Cesarotti, K.~Fraser, S.~Homiller and A.~Parikh,
Phys. Rev. D \textbf{108}, no.5, 055026 (2023)
doi:10.1103/PhysRevD.108.055026
[arXiv:2204.05283 [hep-ph]].

\bibitem{Klimenko:1984qx}
K.~G.~Klimenko,
Theor. Math. Phys. \textbf{62}, 58-65 (1985)
doi:10.1007/BF01034825

\end{thebibliography}
\end{document}